\documentclass[12pt]{iopart}
\usepackage{iopams}
\usepackage{setstack}
\usepackage[dvips]{graphicx}
\usepackage{epsfig}

\begin{document}

\title{Dynamics of continuous-time quantum walks in restricted geometries}

\author{E. Agliari, A. Blumen and O. M\"{u}lken}
\address{Theoretische Polymerphysik, Universit\"{a}t Freiburg,
Hermann-Herder-Str. 3, D-79104 Freiburg, Germany}

\begin{abstract}
We study quantum transport on finite discrete structures and we model the process by means of continuous-time quantum walks.
A direct and effective comparison between quantum and classical walks can be attained based on the average displacement of the walker as a function of time. Indeed, a fast growth of the average displacement can be advantageously exploited to build up efficient search algorithms. By means of analytical and numerical investigations, we show that the finiteness and the inhomogeneity of the substrate jointly weaken the quantum walk performance.
We further highlight the interplay between the quantum-walk dynamics and the underlying topology by studying the temporal evolution of the transfer probability distribution and the lower bound of long time averages.
\end{abstract}

\pacs{05.60Gg, 71.35.-y, 05.60Cd}% PACS, the Physics and Astronomy
\maketitle

\section{Introduction} \label{sec:intro}
Quantum walks (QWs) are attracting increasing attention in many research areas, ranging from solid-state physics to quantum computing \cite{kempe}. In particular, QWs provide a model for quantum-mechanical transport processes on discrete structures; this includes, for instance, the coherent energy transfer of a qubit on an optical lattice \cite{sanders,pozzi,dur,cote}.
The theoretical study of QWs is also encouraged by recent experimental implementations able to corroborate theoretical findings \cite{zou,ryan,blumen_lett}.

As in the classical random walk, quantum walks appear in a discrete \cite{aharonov} as in a
continuous-time (CTQW) \cite{farhi} form; these forms, however, cannot be simply related to each other \cite{strauch}. Now, standard CTQWs, on which we focus, can be obtained by identifying the Hamiltonian of the system with the classical transfer matrix which is, in turn, directly related to the Laplacian of the underlying structure.

Another feature which CTQWs share with classical random walks
consists in the strong interplay between the dynamics properties
displayed by the walk and the topology of the substrate
\cite{jex,volta,pernice}. However, the dependencies turn out to
be much more complex in the quantum-mechanical case: while the
classical (simple) walk eventually loses memory of its starting
site, the quantum walk exhibits, even in the asymptotic regime,
transition probabilities which depend on the starting site. For
this reason, often the parameters describing the transport are
averaged over all initial sites, a procedure which allows a global
characterization of the walk, while preserving its most important
features.

One of the quantities affected by topology is the mean square
displacement of the walker up to time $t$. Classically, this
quantity is monotonically increasing and depends (asymptotically)
on time according to the power law $\langle r^2(t) \rangle \sim
t^{\beta}$. The value of the ``diffusion exponent'' $\beta$ allows
to distinguish between normal ($\beta = 1$) and anomalous ($\beta
\neq 1$) diffusion \cite{havlin}. As for quantum transport, it is possible to
introduce analogous exponents, characterizing the temporal
spreading of a wave-packet \cite{vidal}. However, even when they
take place over the same structure, quantum and classical walks
can exhibit dramatically different behaviours. In particular, the
quantum wave propagation on regular, infinite lattices is
ballistic, i.e. the root mean square displacement is linear in
time. Such a quadratic speed-up of the mean square displacement is
a well known phenomenon when dealing with tight-binding electron
waves on periodic lattices \cite{aharonov} and, from a
computational point of view, it constitutes an important feature
since it could be advantageously exploited in quantum search
algorithms \cite{williams,ambainis,search}.
In the presence of disorder (either deterministic or stochastic)
or finiteness, the sharp ballistic fronts are softened, a fact
which may even lead to the localization of the quantum particle
\cite{yin,yuan}.
It is therefore of both theoretical and practical interest to
highlight how finiteness and inhomogeneity - often unavoidable in
real systems - affect the particle's propagation. To this aim we
analyze quantum transport on restricted geometries, where the
restrictions arise from the (possibly joint) fractal dimension and
finite extent of the substrate itself. By direct comparison with the classical case, we find that, on finite substrates, the advantage of CTQWs is at short times only. Moreover, the lack of
translational invariance weakens the CTQW performance, i.e. in
such situations the average displacement increases more slowly
with time.

The finite discrete structures we consider and compare are the
Dual Sierpinski Gasket (DSG), the Cayley Tree (CT) \textit{vide infra} Sec.~\ref{sec:prob}, and the square
lattice with periodic boundary conditions, i.e. the square torus
(ST). These constitute representative topologies, providing examples of
fractals with loops, of trees and of regular structures. The Dual Sierpinski
Gasket will be treated in more detail; for this structure the eigenvalue
spectrum of the Laplacian matrix is known exactly, allowing for some
analytical estimates.
Indeed, not only random walks, but also many dynamical properties of
connected structures themselves (such as the vibrational structures and the
relaxation modes) depend on the spectrum of their Laplacian matrix
\cite{mohar}. However, for CTQWs the set of eigenvectors also matters,
which often makes analytical investigations cumbersome.

It is worth underlining that focusing on discrete structures
is not only suggested by solid-state applications: quantum
computation is traditionally concerned with the manipulation of
discrete systems. In particular, a discrete (and finite) state
space makes the CTQW simulation by quantum computers, working with
discrete registers, feasible \cite{kempe,vincenzo}.

Our paper is structured as follows. After a brief summary of the main concepts and of the formul{\ae} concerning CTQWs in Sec.~\ref{sec:CTRW}, we describe the topology of the DSG in Sec.~\ref{sec:DSG}. Then, in Sec.~\ref{sec:prob}, we study the quantum-mechanical transport over the above mentioned structures, especially focusing on the average displacement
and on the long time averages. Finally, in Sec.~\ref{sec:concl} we present our comments and conclusions. In the Appendix we derive analytical results concerning the average chemical displacements of CTQWs over hypercubic lattices, special cases being chains and square lattices.

\section{Continuous-time quantum walks on graphs}
\label{sec:CTRW}Mathematically, a graph is specified by the pair $\{V,E\}$
consisting of a nonempty, countable set of points $V$, joined
pairwise by a set of links $E$. The cardinality of $V$ provides
the number $\mathcal{N}$ of sites making up the graph, i.e. its
volume: $|V|=\mathcal{N}$. In the following, we focus mainly on finite
graphs ($\mathcal{N}< \infty$) and we label each node with a lowercase
letter $i \in V$.

From an algebraic point of view, a graph can be described by its
adjacency matrix $\mathbf{A}$, whose elements are
\begin{displaymath}
A_{ij}=
\left\{
\begin{array}{rl}
1 & \mbox{if } (i,j) \in E  \\
0 & \mbox{otherwise}.
\end{array}
\right.
\end{displaymath}
The connectivity of a node $i$ can be calculated as a sum of
matrix elements $z_i = \sum_j A_{ij}$. The Laplacian operator is
then defined as $\mathbf{L} = \mathbf{Z} - \mathbf{A}$, where $\mathbf{Z}$
is the diagonal matrix given by $Z_{ik}=z_i \delta_{ik}$.

The Laplacian matrix $\mathbf{L}$ is symmetric and non-negative
definite and it can therefore generate a probability conserving
Markov process and define a unitary process as well. Otherwise
stated, the Laplacian operator can work both as a classical
transfer operator and as a tight-binding Hamiltonian of a quantum
transport process \cite{childs,volta2}.

Indeed, the classical continuous-time random walk (CTRW) is described by the following Master equation
\cite{weiss}:
\begin{equation}\label{eq:master_cl}
\frac{d}{dt} p_{k,j}(t)= \sum_{l=1}^{\mathcal{N}} T_{kl} p_{l,j}(t),
\end{equation}
where $p_{k,j}(t)$ is the conditional probability that the walker
is on node $k$ when it started from node $j$ at time $0$. If the
walk is symmetric with a  site-independent transmission rate $\gamma$, then the
transfer matrix $\mathbf{T}$ is simply related to the Laplacian
operator through $\mathbf{T} = - \gamma \mathbf{L}$.

Now the CTQW, the quantum-mechanical counterpart of the CTRW, is
introduced by identifying the Hamiltonian of the system with the
classical transfer matrix, $\mathbf{H}=-\mathbf{T}$
\cite{farhi,volta,childs} (in the following we will set
$\hbar\equiv1$).
The set of states $|j \rangle$, representing the walker localized
at the node $j$, spans the whole accessible
Hilbert space and also provides an orthonormal basis set.
Therefore, the behaviour of the walker can be described by the transition
amplitude $\alpha_{k,j}(t)$ from state $| j \rangle$ to
state $| k \rangle$, which obeys the following
Schr\"{o}dinger equation:
\begin{equation}\label{eq:schrodinger}
\frac{d}{dt} \alpha_{k,j}(t)=-i \sum_{l=1}^{\mathcal{N}} H_{kl} \alpha_{l,j}(t).
\end{equation}
If at the initial time $t_0=0$ only the state $| j \rangle$ is
populated, then the formal solution to Eq.~\ref{eq:schrodinger} can be
written as
\begin{equation}\label{eq:formal_solution}
\alpha_{k,j}(t)=\langle k | \exp(-i\mathbf{H}t)   | j
\rangle,
\end{equation}
whose squared magnitude provides the quantum mechanical
transition probability $\pi_{k,j}(t) \equiv |\alpha_{k,j}(t)|^2$.
In general, it is convenient to introduce the orthonormal basis $|
\psi_n \rangle, n \in [1,\mathcal{N}]$ which diagonalizes
$\mathbf{T}$ (and, clearly, also $\mathbf{H}$); the correspondent
set of eigenvalues is denoted by $\{\lambda_n\}_{n=1,...,\mathcal{N}}$. Thus, we can
write
\begin{equation}
\pi_{k,j}(t) = \left| \sum_{n=1}^{\mathcal{N}} \langle k |
e^{-i \lambda_n t} |\psi_n \rangle \langle \psi_n | j \rangle
\right|^2.
\end{equation}

Despite the apparent similarity between Eq.~\ref{eq:master_cl} and
\ref{eq:schrodinger}, some important differences are worth being
recalled.\\
First of all, the imaginary unit makes the time evolution operator
$\mathbf{U}(t)=\exp(-i \mathbf{H} t)$ unitary, which prevents the
quantum mechanical transition probability from having a definite
limit as $t \rightarrow \infty$. On the other hand, a particle
performing a CTRW is asymptotically equally likely to be found on
any site of the structure: the classical $p_{k,j}(t)$ admit a
stationary distribution which is independent of initial and final
sites, $\lim_{t \rightarrow \infty}p_{k,j}(t)
=1 / \mathcal{N}$. Hence, in order to compare classical
long time probabilities with quantum mechanical ones, we rely on
the long time average (LTA)
\cite{aharonov2}, defined in Sec.~\ref{ssec:LongTime}.\\
Moreover, the normalization conditions for $p_{k,j}(t)$ and
$\alpha_{k,j}(t)$ read $\sum_{k=1}^{\mathcal{N}} p_{k,j}(t)=1$,
and $\sum_{k=1}^{\mathcal{N}} |\alpha_{k,j}(t)|^2=1.$

\subsection{Average displacement}\label{ssec:AD}
The average displacement performed by a quantum
walker until time $t$ allows a straightforward comparison with
the classical case; it is also more directly related to
transport properties than the transfer probability $\pi_{k,j}(t)$:
It constitutes the expectation value of the distance reached by
the particle after a time $t$ and its time dependence provides
information on how fast the particle propagates over the
substrate.

For CTQW (subscript $q$) starting at node $j$, we define the \textit{average (chemical)
displacement} $\langle r_j(t) \rangle_q$ performed until time $t$
as
\begin{equation} \label{eq:exploration_depth}
\langle r_j(t)\rangle_q = \sum_{k=1}^{\mathcal{N}}  \ell(k,j)
\pi_{k,j}(t),
\end{equation}
where $\ell(k,j)$ is the chemical distance between the sites $j$
and $k$, i.e. the length of the shortest path connecting $j$ and
$k$. We can average over all starting points to obtain
\begin{equation} \label{eq:av_cover_length}
\overline{\langle r(t)\rangle}_q = \frac{1}{\mathcal{N}} \sum_{j=1}^{\mathcal{N}} \langle
r_j(t)\rangle_q.
\end{equation}

For fractals or hyperbranched structures it is more appropriate to
use the chemical distance, rather than the Euclidean distance; for
instance, the infinite CT (see Sec.~\ref{sec:prob}) cannot be embedded in any lattice of
finite dimension. For classical diffusion, it is well-known that
the chemical and the Euclidean distances display analogous asymptotic
laws for regular structures and for many deterministic fractals
(e.g. the Sierpinski Gasket) \cite{havlin}; as discussed in the Appendix, this still holds for CTQWs on arbitrary $d$-dimensional hypercubic lattices.

For classical (subscript $c$) regular diffusion (on infinite lattices) the
average displacement $\langle r(t) \rangle_c$ depends on time $t$
according to
\begin{equation}\label{eq:normal_diffusion}
\langle r(t) \rangle_c \sim t^{1/2}.
\end{equation}
More generally, for scaling (fractal) structures
%characterized by the
%fractal dimension $d_f$ and the spectral dimension $\tilde{d}$,
we can define the
so-called chemical diffusion exponent $d_w^{\ell}$ and get \cite{havlin}:
\begin{equation}\label{eq:anomalous_diffusion}
\langle r(t) \rangle_c \sim t^{1/d_w^{\ell}}.
\end{equation}
%In particular, for the DSG $d_w=\ln 5 / \ln 3 \simeq 2.3219$,
%whereas,
%of course, for translationally invariant structures, $d_w^{\ell}=2$, so that
%Eq.~\ref{eq:normal_diffusion} is recovered.
Finite systems require
corrections to these laws: for them $\langle r(t) \rangle_c$
does not grow indefinitely, but it saturates to a maximum value
$r_c$ \cite{reis}.

\subsection{Return Probability}
As it is well known, for a diffusive particle the probability to return
to the starting point is topology sensitive, and it can indeed
be used to extract information about the underlying structure
\cite{weiss}. It is therefore interesting to compare the classical
return probability $p_{k,k}(t)$ with the quantum mechanical
$\pi_{k,k}(t)$ (see also \cite{muelken,muelken3}). One has
\begin{equation}
p_{k,k}(t)=\langle k | \exp(\mathbf{T} t) | k \rangle = \sum_{n=1}^{\mathcal{N}}
\left|\langle k | \psi_n \rangle \right|^2 \exp(- \gamma t
\lambda_n)
\end{equation}
and
\begin{equation}
\pi_{k,k}(t)= \left| \alpha_{k,k}(t) \right|^2 = \left| \sum_{n=1}^{\mathcal{N}}
\left|\langle k | \psi_n \rangle \right|^2 \exp(- i \gamma t
\lambda_n) \right|^2 \label{eq:pi_kk}.
\end{equation}
In order to get a global information about the likelihood to be (return or
stay) at the origin, independent of the starting site, we average over all sites of the graph, obtaining
\begin{equation} \label{eq:p_bar}
\bar{p}(t) = \frac{1}{\mathcal{N}} \sum_{k=1}^{\mathcal{N}}
p_{k,k}(t)=\frac{1}{\mathcal{N}} \sum_{n=1}^{\mathcal{N}} e^{-\gamma \lambda_n t}
\end{equation}
and
\begin{equation}\label{eq:pi_bar}
\bar{\pi}(t) = \frac{1}{\mathcal{N}} \sum_{k=1}^{\mathcal{N}} \pi_{k,k}(t) = \frac{1}{\mathcal{N}} \sum_{n,m=1}^{\mathcal{N}} e^{-i \gamma (\lambda_n -
\lambda_m)t} \sum_{k=1}^{\mathcal{N}}  \left|\langle k | \psi_n \rangle \right|^2
\left|\langle k | \psi_m \rangle \right|^2.
\end{equation}

For finite substrates,
the classical $\bar{p}(t)$ decays monotonically
to the equipartition limit, and it only
depends on the eigenvalues of $\mathbf{T}$. On the other hand, $\bar{\pi}(t)$
depends explicitly on the eigenvectors of $\mathbf{H}$ \cite{muelken,muelken3}.
By means of the Cauchy-Schwarz inequality we can obtain a lower bound for
$\bar{\pi}(t)$ which does not depend on the eigenvectors
\cite{muelken3,bierbaum}:
\begin{equation} \label{eq:mu}
\bar{\pi}(t) \geq \left| \frac{1}{\mathcal{N}} \sum_{k=1}^{\mathcal{N}}
\alpha_{k,k}(t) \right| \equiv |\bar{\alpha}(t)|^2 = \frac{1}{\mathcal{N}^2} \sum_{m,n=1}^{\mathcal{N}} e^ {-i
\gamma (\lambda_n - \lambda_m)t  }.% = \mu(t).
\end{equation}
Notice that Eqs.~\ref{eq:p_bar} and \ref{eq:pi_bar} can serve as
measures of the efficiency of the transport process performed by
CTRW and CTQW, respectively. In fact, the faster $\bar{p}(t)$
decreases towards its asymptotic value, the more efficient the
transport.  Analogously, a more rapid decay of the envelope of
$\bar{\pi}(t)$ (or of $|\bar{\alpha}(t)|^2$) implies a faster
delocalization of the quantum walker over the graph. By the way, we recall
that, for a large variety of graphs \cite{muelken3}, the classical average
return probability scales as $\bar{p}(t) \sim t^{-\mu}$, while the
envelope of $|\bar{\alpha}(t)|^2$, namely $\mathrm{env} \left[
|\bar{\alpha}(t)|^2 \right]$, scales like $t^{-2 \mu}$, $\mu$ being a proper
parameter related for fractals to the spectral density.

As can be inferred by comparing Eqs.~\ref{eq:p_bar} and
\ref{eq:pi_bar}, for quantum transport processes the degeneracy of
the eigenvalues plays an important role, as the differences
between eigenvalues determine the temporal behaviour, while for
classical transport the long time behaviour is dominated by the
smallest eigenvalue. Situations in which only a few, highly
degenerate eigenvalues are present are related to slow CTQW
dynamics, while when all eigenvalues are non-degenerate the
transport turns out to be efficient \cite{muelken,muelken3}.

\section{Dual Sierpinski gasket: Topology and eigenvalue spectrum}\label{sec:DSG}
\begin{figure}
\begin{center}
\resizebox{0.85\columnwidth}{!}{\includegraphics{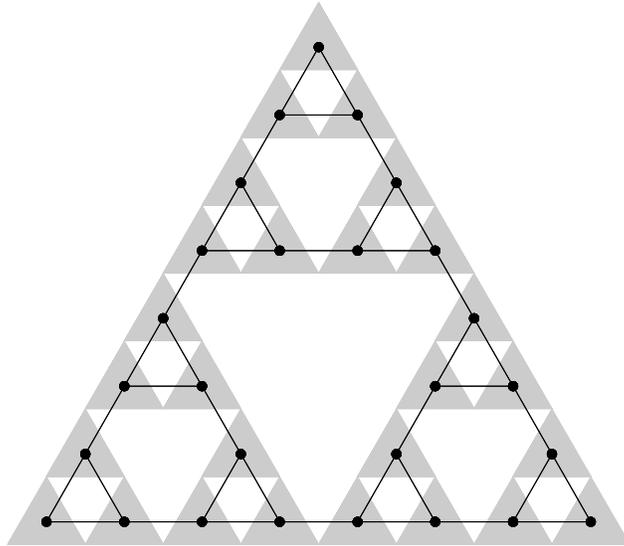}}
\end{center}
\caption{Dual transformation from Sierpinski Gasket to Dual Sierpinski Gasket of generation $g=3$.} \label{fig:DualTransf}
\end{figure}
Before turning to the dynamics of CTQW (and CTRW) on exemplary
structures, which allow us to highlight the importance of
inhomogeneities, some remarks on the spectra of the DSG are in
order.

The dual Sierpinski gasket is an exactly-decimable fractal which
is directly related, through a dual transformation, to the
Sierpinski gasket (SG). The DSG of generation $g$ can be
constructed by replacing each small triangle belonging
to the SG with a node and by connecting such nodes whenever the relevant
triangles share a vertex in the original gasket (see
Fig.~\ref{fig:DualTransf}). It is straightforward to verify that
the number of nodes at any given generation $g$ is
$\mathcal{N}=3^g$.

The dual transformation does not conserve the coordination number of the inner nodes
(which decreases from 4 to 3), but it does conserve the fractal
dimension $d_f$ and the spectral dimension $\tilde{d}$, which are
therefore the same as for the original Sierpinski gasket $d_f =
\ln 3 / \ln 2 = 1.58496...$ and $\tilde{d}= 2 \ln 3 / \ln
5=1.36521...$.

As mentioned above, the knowledge of the eigenvalue spectrum is
sufficient for the calculation of several interesting quantities
concerning the dynamics of CTQWs. In general, any (finite)
Hamiltonian $\mathbf{H}$ can be (at least numerically)
diagonalized in order to obtain its spectrum. However, as the size
of $\mathbf{H}$ gets large, the procedure gets to be time
consuming and the precise numerical diagonalization may not be
easy to perform. Remarkably, the eigenvalue spectrum of the DSG
Laplacian matrix can be determined at any generation through the
following iterative procedure; for more details we refer to
\cite{cosenza,blumen}: At any given generation $g$ the spectrum
includes the non-degenerate eigenvalue $\lambda_{\mathcal{N}}=0$,
the eigenvalue $3$ with degeneracy $(3^{g-1} + 3)/2$ and the
eigenvalue $5$ with degeneracy $(3^{g-1} - 1)/2$. Moreover, given
the eigenvalue spectrum at generation $g-1$, each
non-vanishing eigenvalue $\lambda_{g-1}$ corresponds to two new
eigenvalues $\lambda_g^{\pm}$ according to
\begin{equation}
\lambda_g^{\pm}=\frac{5 \pm \sqrt{25-4 \lambda_{g-1}}}{2};
\end{equation}
both $\lambda_{g}^+$ and $\lambda_{g}^-$ inherit the degeneracy of
$\lambda_{g-1}$. The eigenvalue spectra is therefore bounded in
$[0,5]$. As explained in \cite{cosenza}, at any generation $g$, we
can calculate the degeneracy of each distinct eigenvalue: apart
from $\lambda_{\mathcal{N}}$ whose degeneracy is $1$, there are
$2^r$ distinct eigenvalues, each with degeneracy
$(3^{g-r-1}+3)/2$, being $r=0,1,...,g-1$, and $2^r$ distinct
eigenvalues, each with degeneracy $(3^{g-r-1}-1)/2$, being
$r=0,1,...,g-2$. As can be easily verified, the degeneracies sum
up to $\mathcal{N}=3^g$.
Finally, notice that the distribution of eigenvalues and their
degeneracies are non-uniform and that the spectrum is
multifractal \cite{cosenza}.

\section{CTQWs on restricted geometries} \label{sec:prob}
\subsection{Transfer probability}\label{subsec:TP}
\begin{figure}
\begin{center}
\resizebox{12cm}{10cm}{\includegraphics{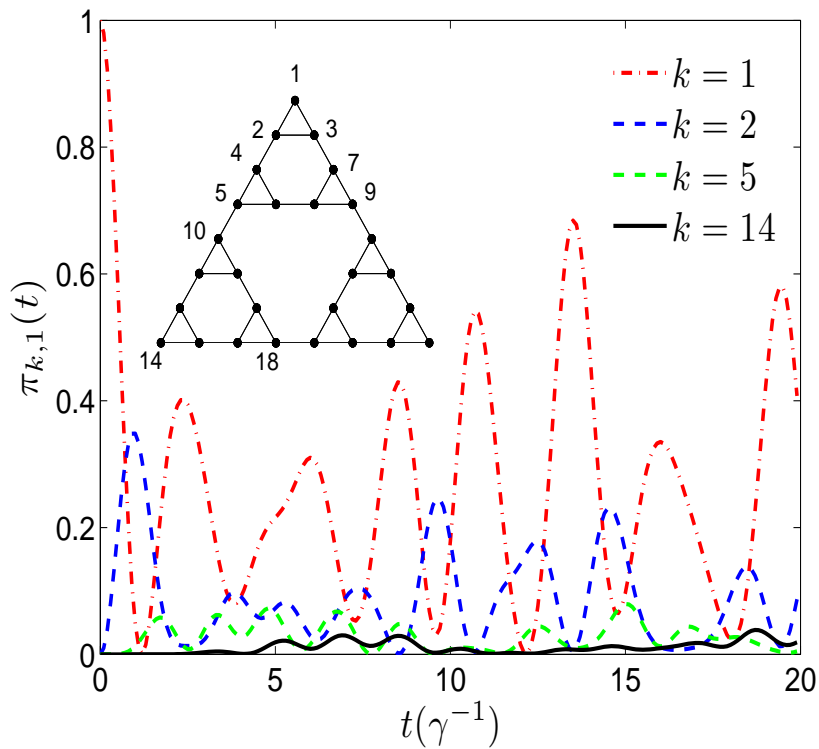}}
\end{center}
\caption{Exact probability $\pi_{k,1}(t)$ for the CTQW starting from
the apex (site $j=1$) to reach sites $k=1,2,5$ and $14$. The $k$-sites are
the left corners at generations $g=0,1,2$ and $3$,
respectively.} \label{fig:pi_apex}
\end{figure}
Results for the exact transition probability distribution $\pi_{k,j}(t)$ for STs and
CTs have already been given in Refs.~\cite{volta,bierbaum,konno1}, where it was shown that $\pi_{k,j}(t)$ depends significantly on the starting node. Results for ultrametric structures are given in \cite{konno2}.

It is worth recalling here that the Cayley tree (CT) can be built by starting from one node (root) connected to $z$ nodes, which constitute the first shell. Each node of the first shell is then connected to $z-1$ new nodes, which constitute the second shell and so forth, iteratively. Therefore, the $M$-th shell contains $z(z-1)^{M-1}$ nodes which are at a chemical distance $M$ from the root. Thus, the CT is a $z$-regular loop-free graph. The numbers of sites in a CT of $M$ shells is $\mathcal{N}_M=[z(z-1)^M-2]/(z-2)$, hence the correlated fractal dimension $\log(\mathcal{N}_M) / \log(M)$ goes to infinity for $M \rightarrow \infty$, precluding the possibility of embedding very large CT in any previously specified Euclidean lattice. In the following we focus on finite $3$-Cayley trees, which means that $z$ is fixed and equal to three for any internal site of the graph; furthermore, the number of shells (also called generation) is finite (and therefore also the number of nodes is itself finite).

In Fig.~\ref{fig:pi_apex} we show our results for a DSG of generation
$g=3$ and we focus on the set of pairs given by
$(v_n,1)$, where $v_n$ denotes any of the two corners of the gasket of the $n$-th generation, with $n
\leqslant g$ (i.e., according to the labeling of
Fig.~\ref{fig:pi_apex}, $v_0 = 1, v_1=2, v_2=5, v_3=14$). Now, due to the
symmetry the DSG is endowed with, for CTQWs starting from a given
vertex, say the apex, the left and right corners are equivalent. As expected, $\pi_{k,j}(t)$ does not converge to any
definite value, but it displays oscillations whose amplitudes and
average values get smaller as the distance between the sites $1$
and $v_n$ increases. This suggests, at least when starting from a
main vertex, that the CTQW stays mainly localized at the origin and its
neighbourhood.

We corroborate this by looking at the temporal evolution of
$\pi_{k,j}(t)$ for the DSG and by comparing it to the $\pi_{k,j}(t)$ pertaining to the CT and the
ST of comparable size $\mathcal{N}$.
\begin{figure}
\resizebox{8.1cm}{6cm}{\includegraphics{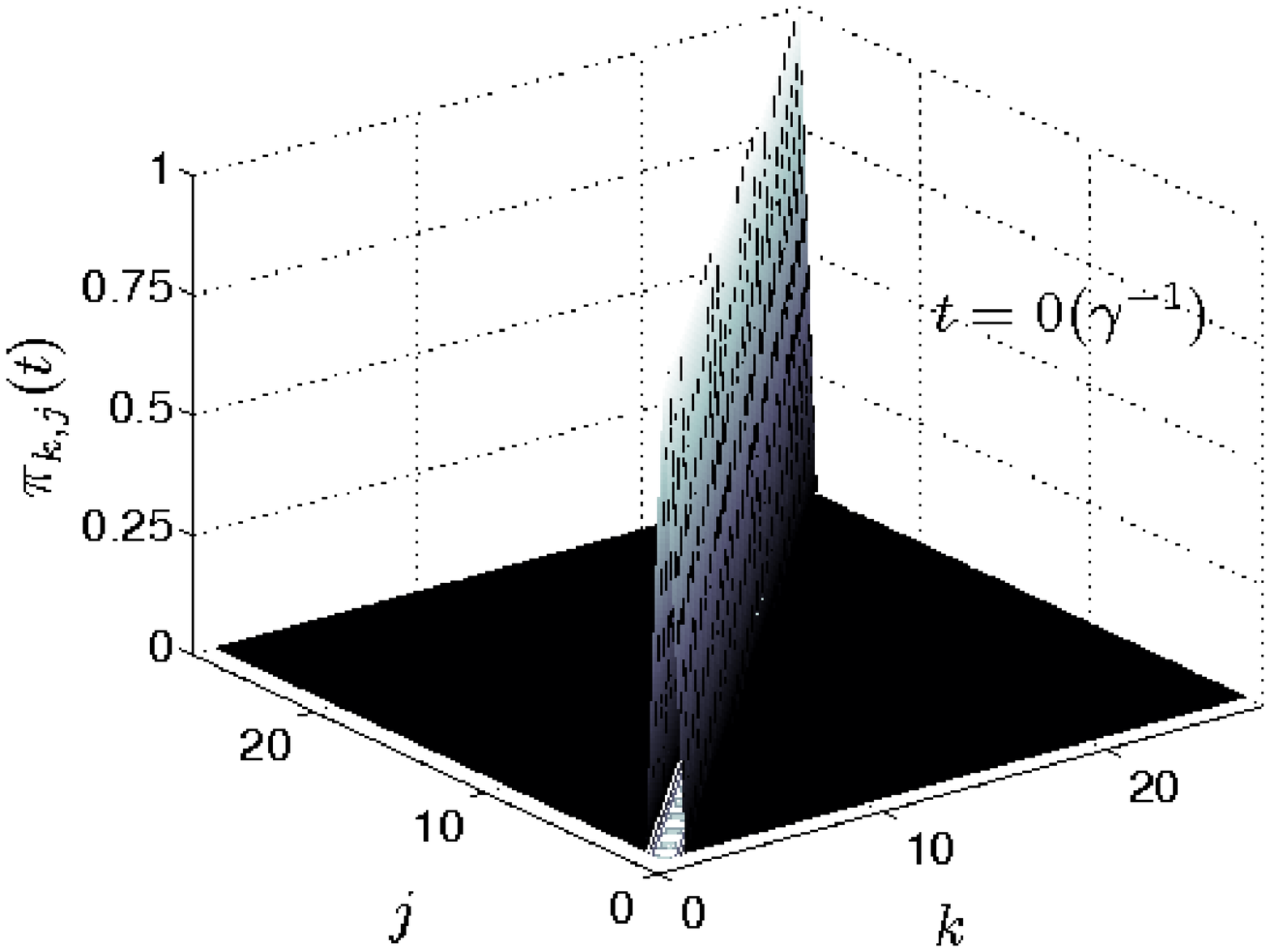}}
\resizebox{8.1cm}{6cm}{\includegraphics{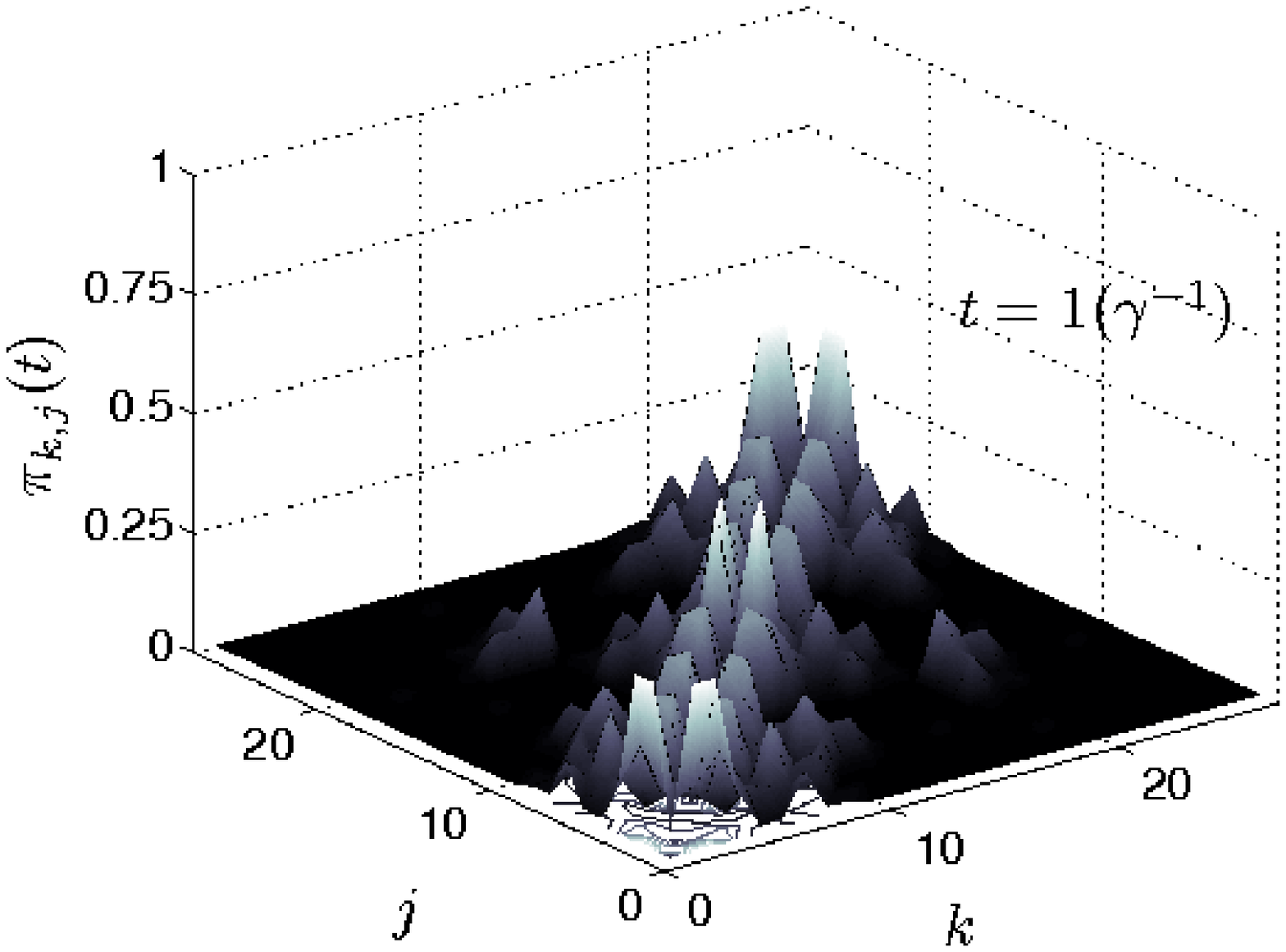}}
\resizebox{8.1cm}{6cm}{\includegraphics{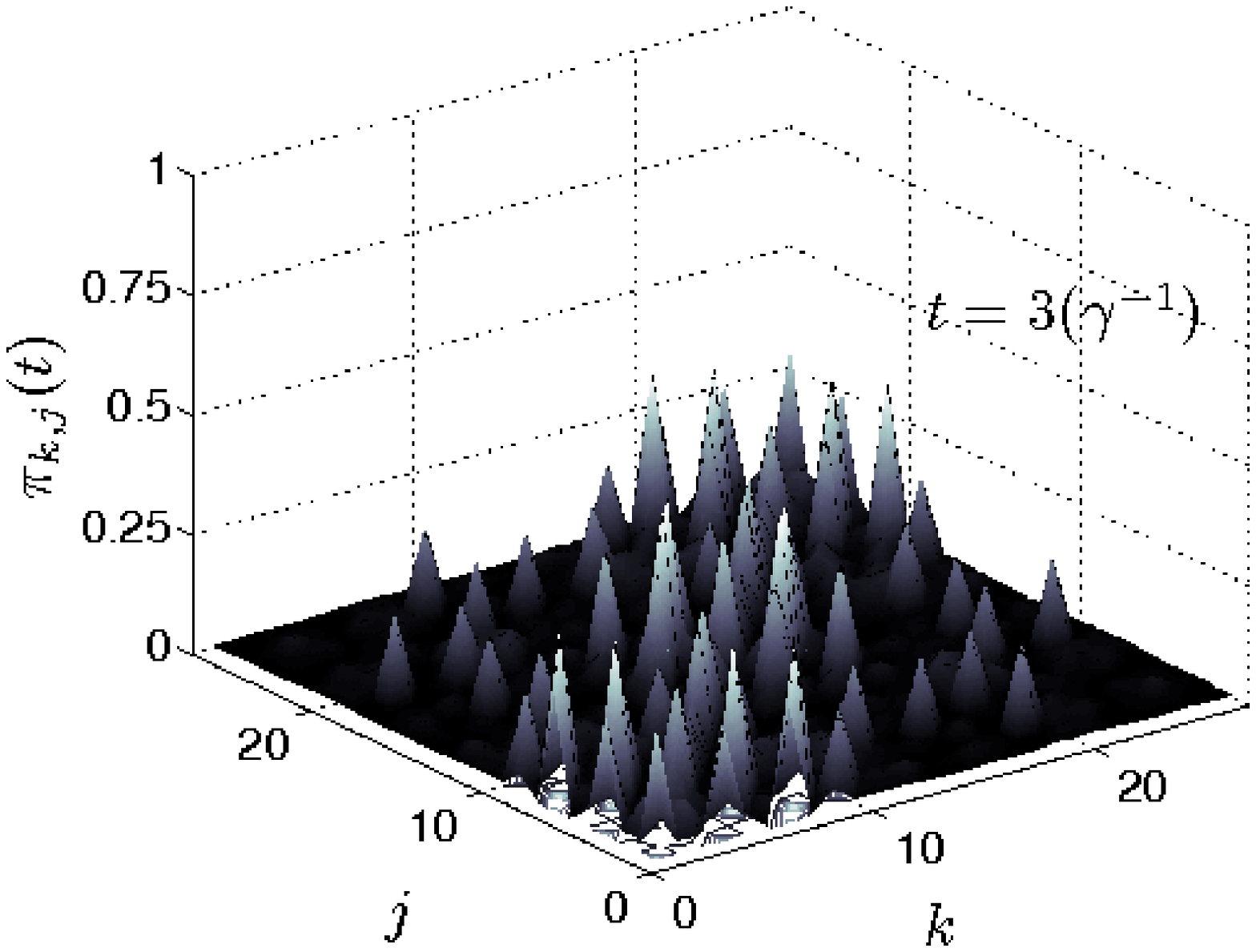}}
\resizebox{8.1cm}{6cm}{\includegraphics{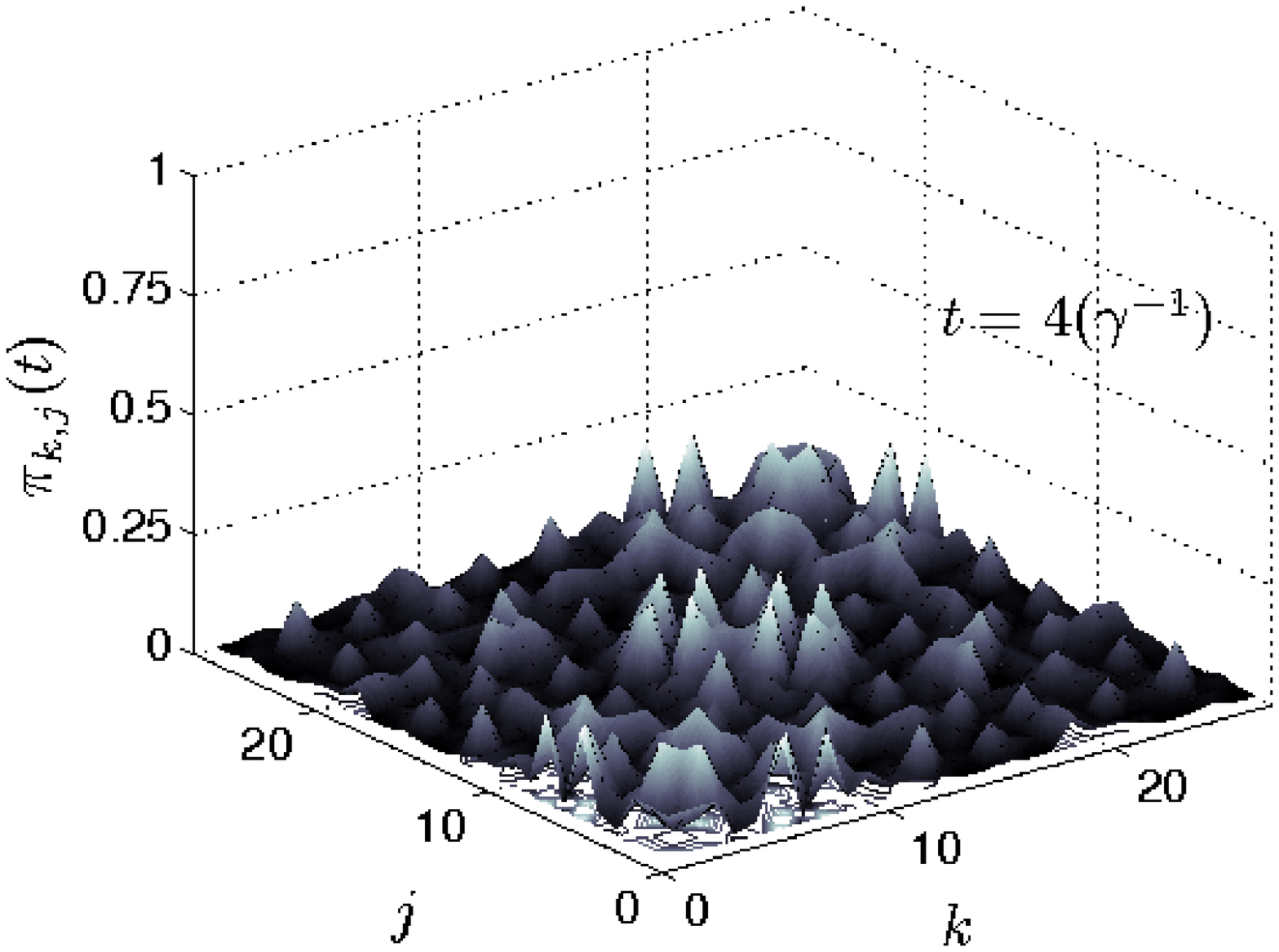}}
\caption{Snapshots of the transition probability $\pi_{k,j}(t)$ at
different times $t=0,1,3,4$ (in units of $\gamma^{-1}$) for the
Dual Sierpinski Gasket of generation $g=3$ ($\mathcal{N}=27$).} \label{fig:movie_DS}
\end{figure}
\begin{figure}
\resizebox{8.1cm}{6cm}{\includegraphics{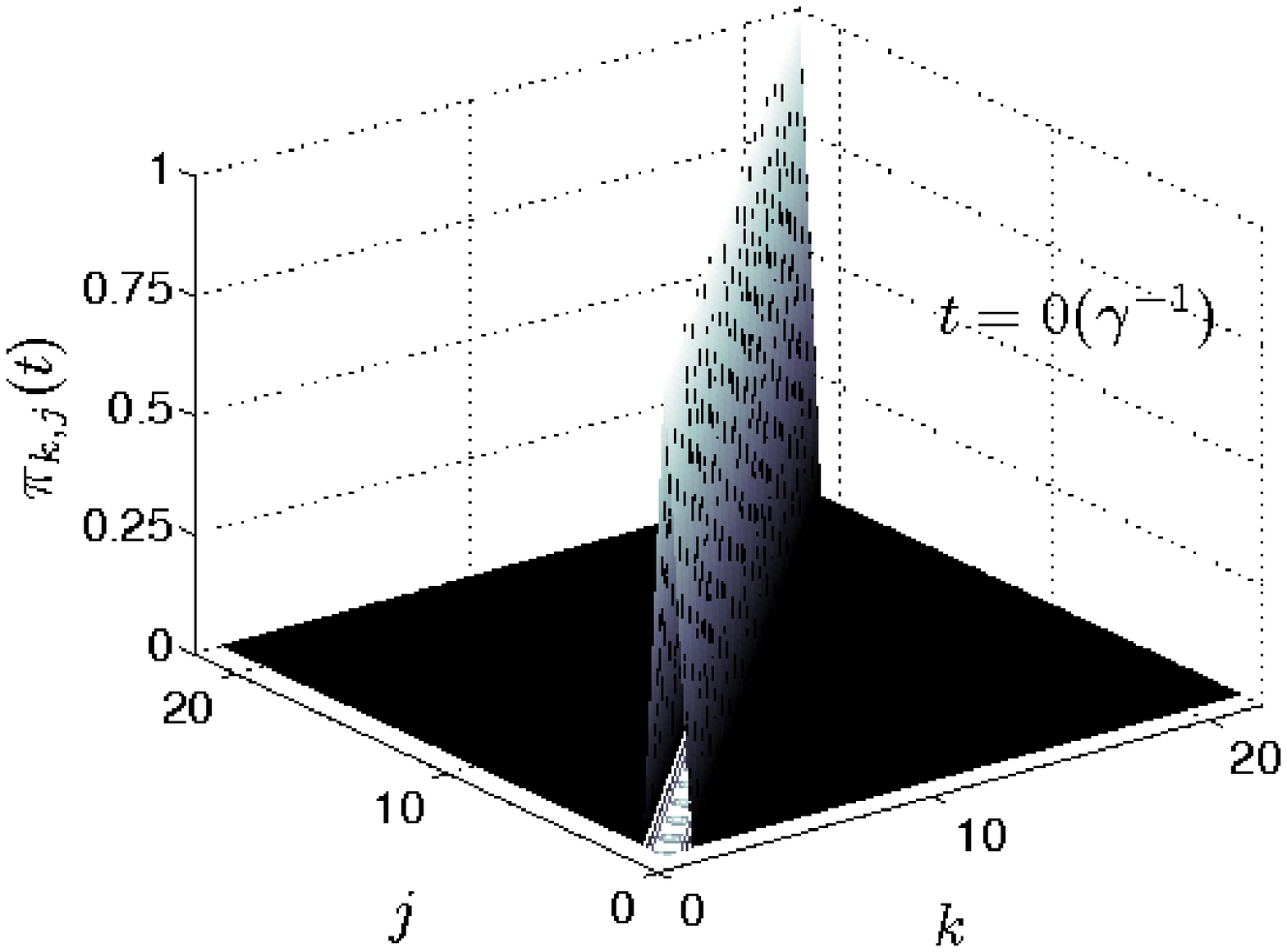}}
\resizebox{8.1cm}{6cm}{\includegraphics{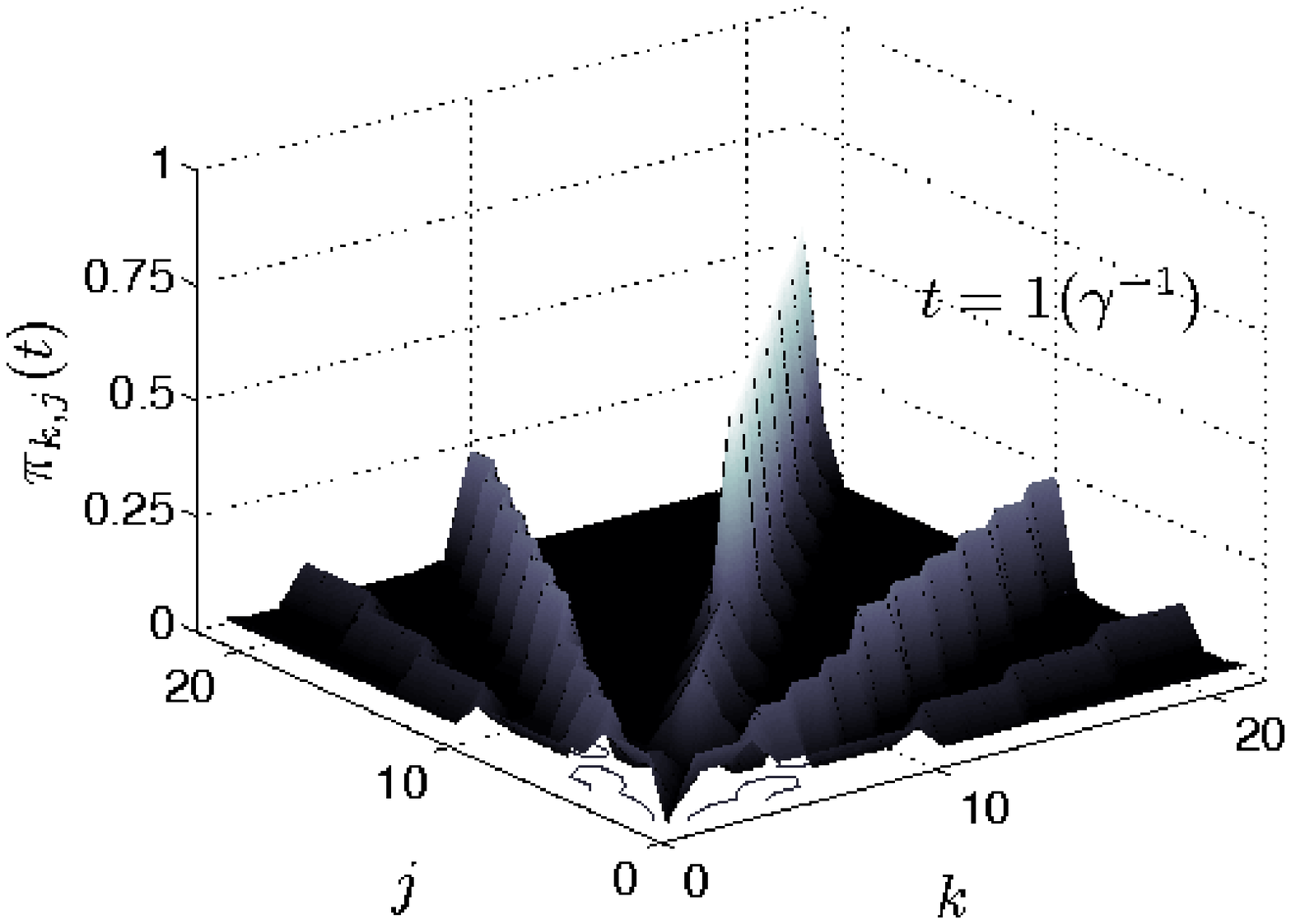}}
\resizebox{8.1cm}{6cm}{\includegraphics{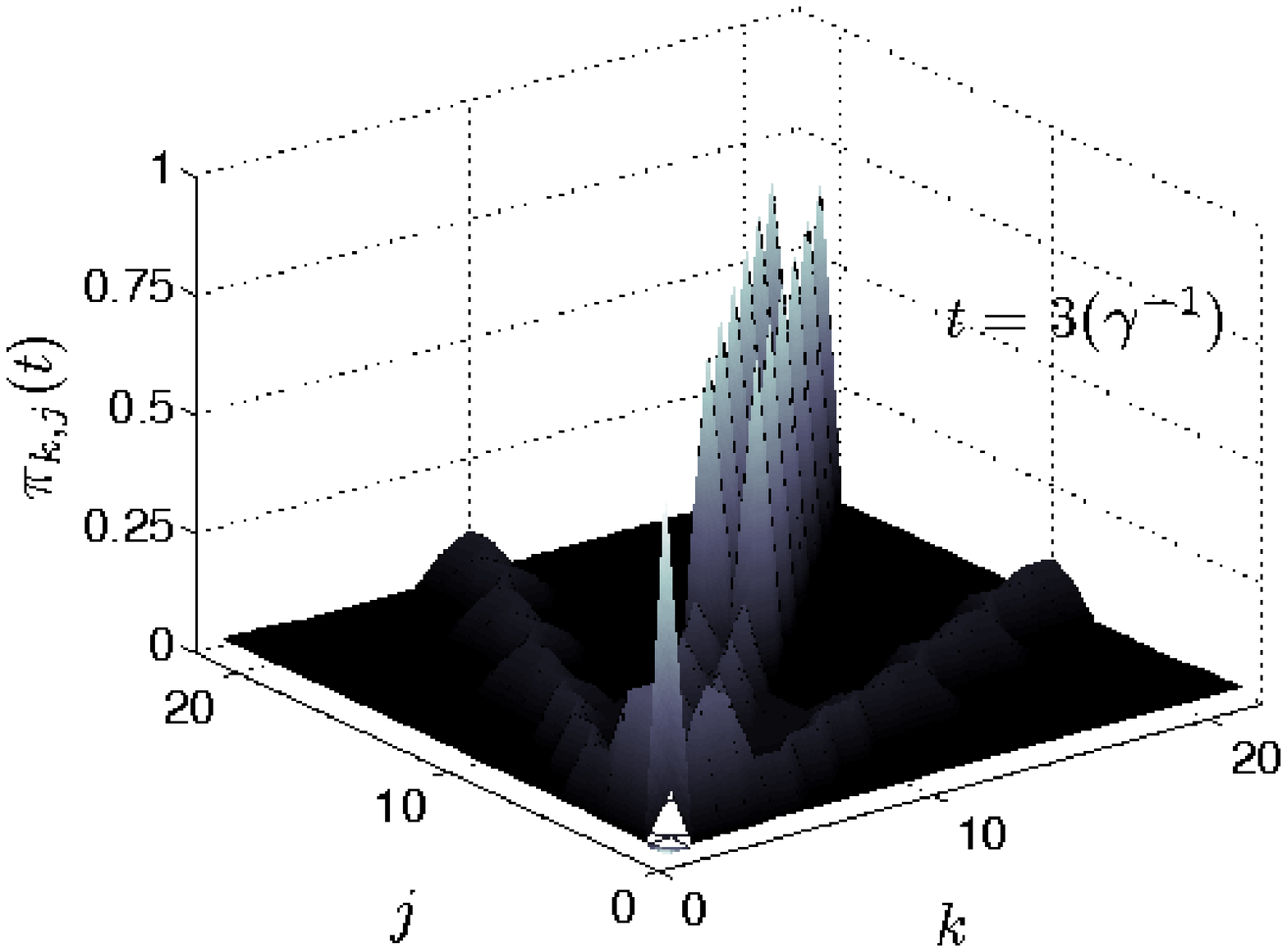}}
\resizebox{8.1cm}{6cm}{\includegraphics{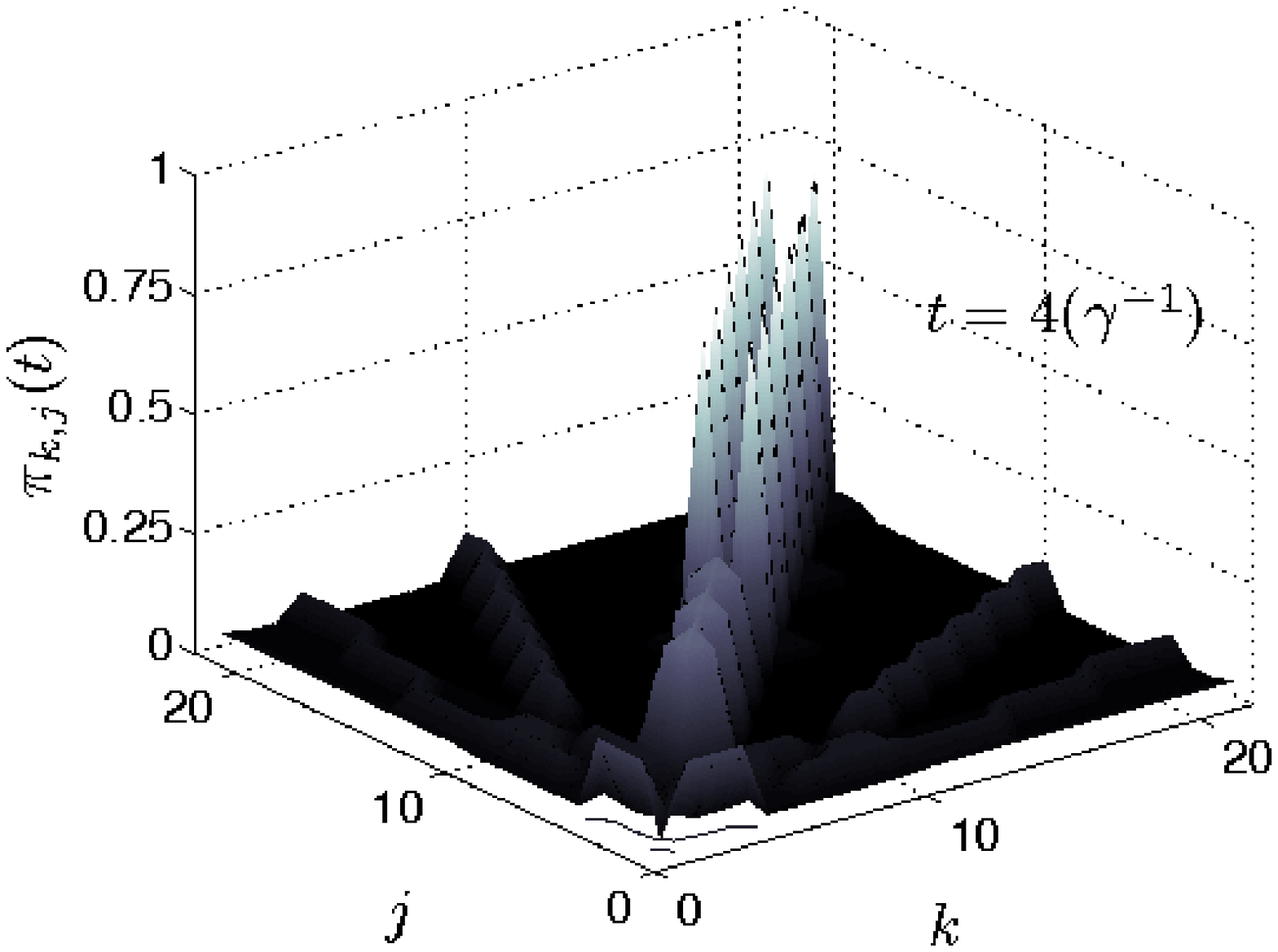}}
\caption{Snapshots of the transition probability $\pi_{k,j}(t)$ at
different times $t=0,1,3,4$ for a Cayley Tree of generation $g=3$ ($\mathcal{N}=3 \times 2^g - 2 = 22$);
time is give in units of ${\gamma}^{-1}$. Notice that the
distribution is localized on special couples of nearest-neighbours
sites and that also reflection effects appears.}
\label{fig:movie_D}
\end{figure}
\begin{figure}
\resizebox{8.1cm}{6cm}{\includegraphics{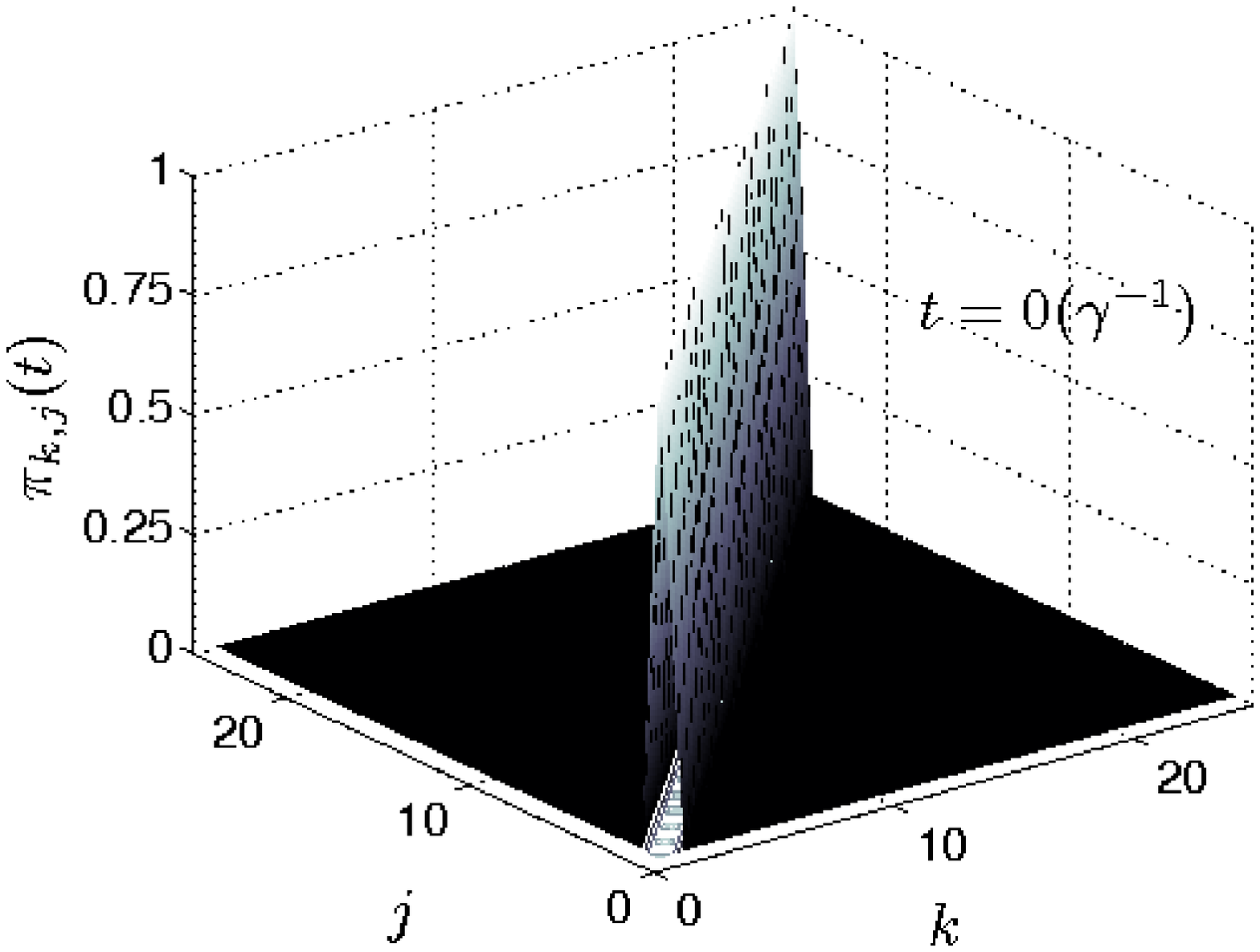}}
\resizebox{8.1cm}{6cm}{\includegraphics{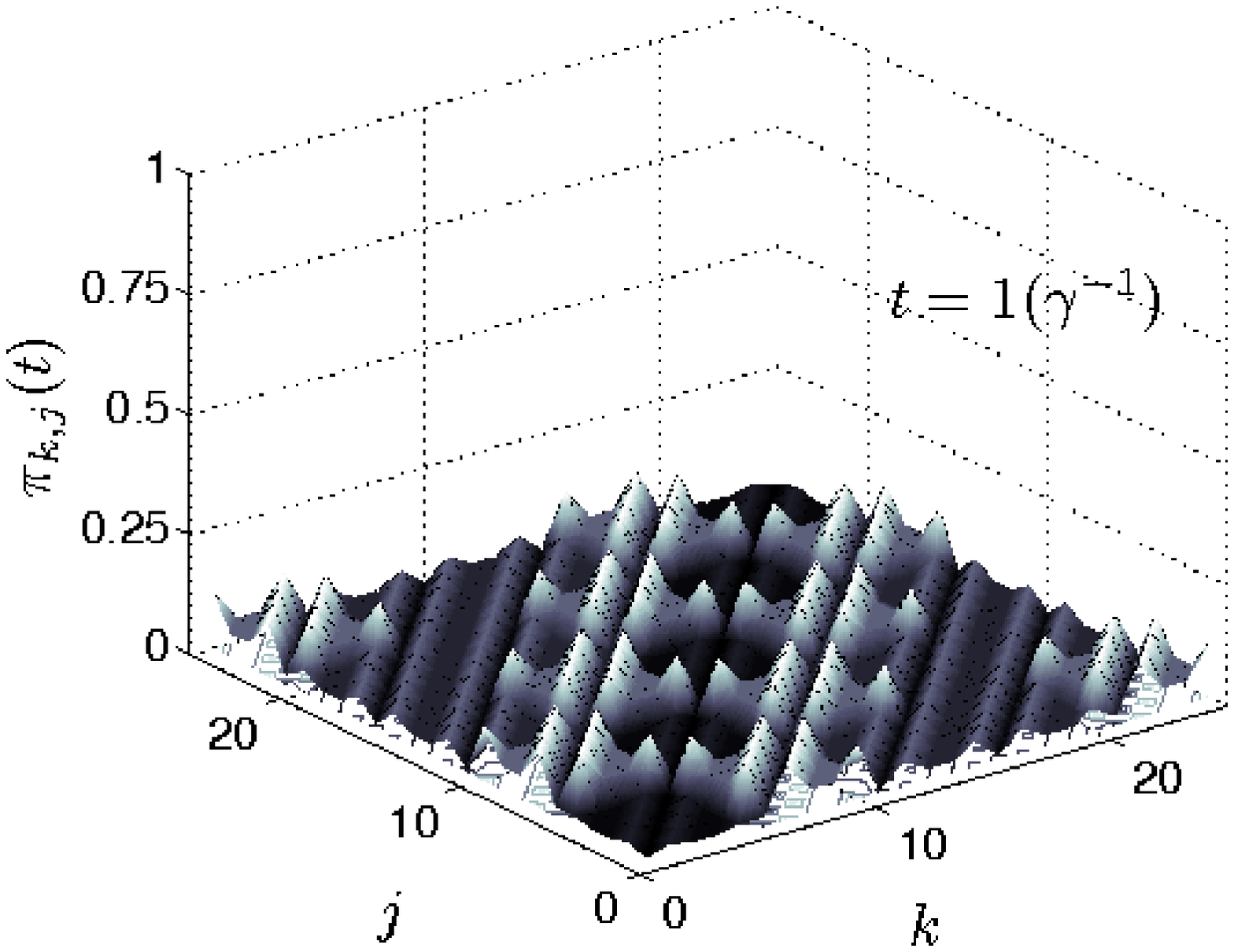}}
\resizebox{8.1cm}{6cm}{\includegraphics{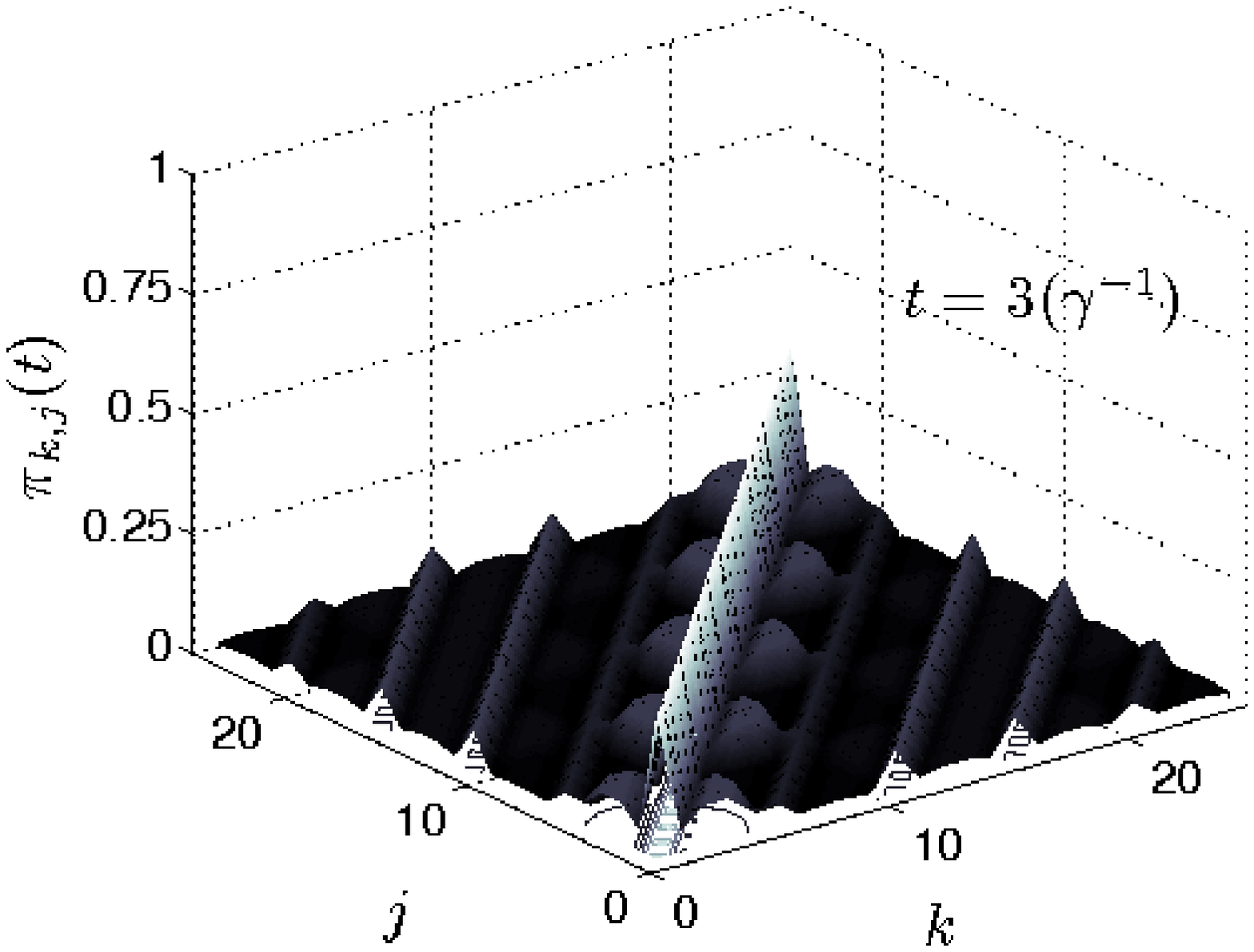}}
\resizebox{8.1cm}{6cm}{\includegraphics{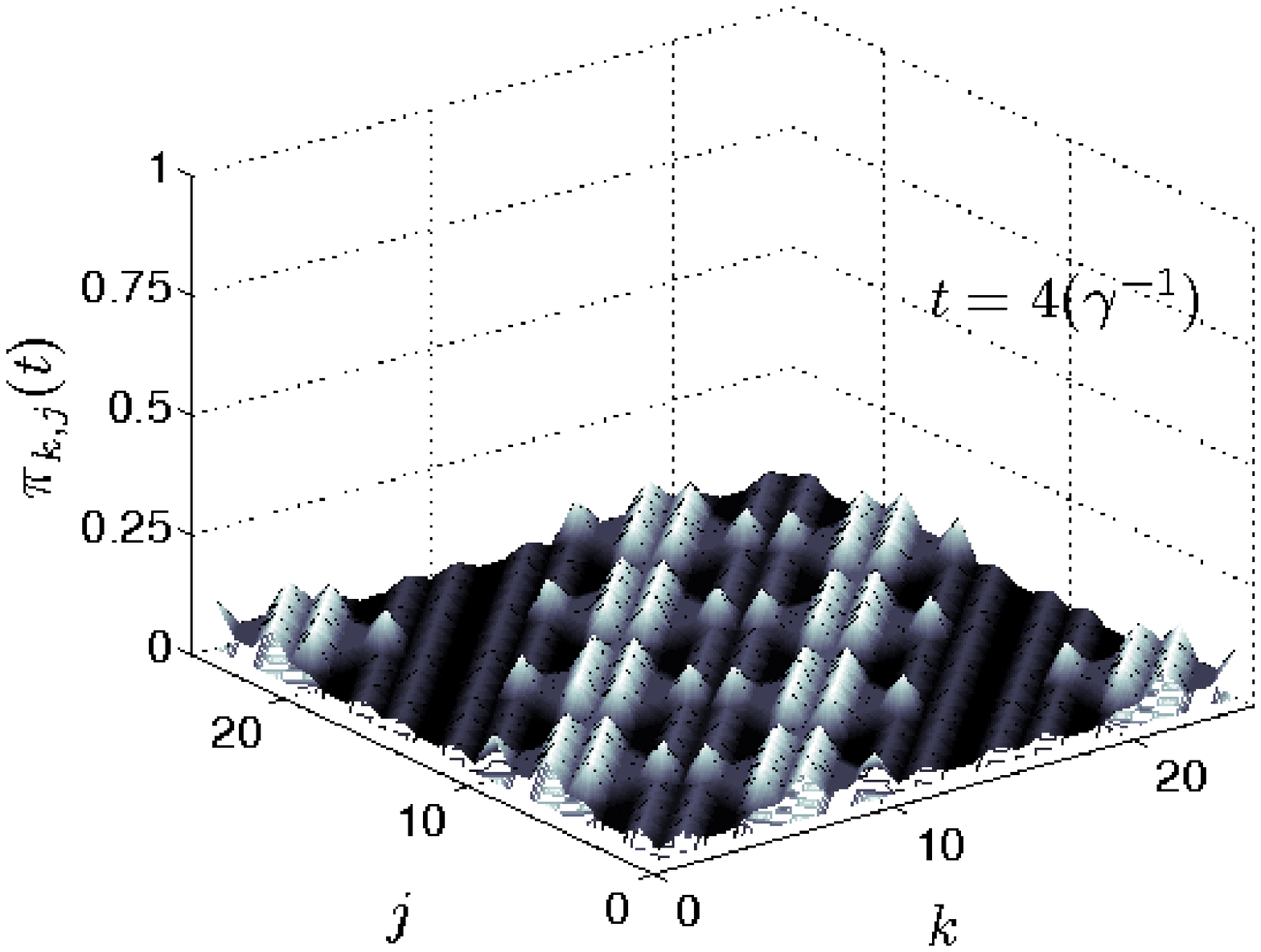}}
\caption{Snapshots of the transition probability $\pi_{k,j}(t)$ at
different times $t=0,1,3,4$ (in units of $\gamma^{-1}$) for the
Square Torus of linear size $L = 5$ $(\mathcal{N}=25)$. Notice that the distribution spreads
very rapidly and regularly over the whole structure.}
\label{fig:movie_Q}
\end{figure}
Figures \ref{fig:movie_DS}, \ref{fig:movie_D} and
\ref{fig:movie_Q} show 3D pictures of $\pi_{k,j}(t)$ at different
moments $t$ (belonging to the short-time regime). On both the $x$ and the
$y$-axis, $k$ and $j$ label the nodes of the graph in such a way
that at the point $(k,j)$ on the $xy$ plane the value of
$\pi_{k,j}(t)$ is presented. At the initial time $t=0$, the
transition probability $\pi_{k,j}(t)$ is non-vanishing only on the
diagonal, i.e. one has $\pi_{k,j}(0)= \delta_{k,j}$; at later times,
$\pi_{k,j}(t)$ spreads out non-uniformly, according to the
topology of the substrate.
% ; the regularity of the substrate is related to the homogeneity of the pattern representing $\pi_{k,j}(t)$.
In particular, for the DSG (Fig.~\ref{fig:movie_DS}), a large fraction of
$\pi_{k,j}(t)$ stays in a region connected by bonds to the initial nodes;
several peaks can be distinguished, whose heights decrease as the chemical
distance between the pertaining sites gets larger. For the CT, the pattern
representing $\pi_{k,j}(t)$ is even more inhomogenous; as can be inferred
from Fig.~\ref{fig:movie_D}, a quantum particle on the CT is located with
very high probability on its initial node (except when starting from the
central node) and, for the time scale considered, it is very unlikely to
reach nodes outside its starting branch.
On the other hand, for the ST (Fig.~\ref{fig:movie_Q}) we notice that the
spread of $\pi_{k,j}(t)$ is rapid and regular: apart from possibly partial
revival phenomena (see for example the snapshot for $t=3 \gamma^{-1}$),
the pattern for $\pi_{k,j}(t)$ exhibits very low peaks.

Indeed, peaks in $\pi_{k,j}(t)$ are a consequence of the
%finiteness of the substrate: one observes
constructive interference stemming from reflections at peripheral sites or
(in the case of the torus) from the superposition of traveling waves which
have crossed the whole (finite) graph.

Finally, Figs. \ref{fig:movie_DS}, \ref{fig:movie_D} and
\ref{fig:movie_Q} also highlight the symmetry characterizing the
quantum transfer probability, namely that
$\pi_{j,k}(t)=\pi_{k,j}(t)$, at all times. This can be derived
directly from Eq.~\ref{eq:formal_solution}, recalling that
$\mathbf{H}$ is itself symmetric and real. An analogous symmetry also characterizes the classical distribution $p_{j,k}(t)$ for all the cases analyzed here.

\subsection{Average displacement}
\begin{figure}
\begin{center}
\resizebox{11.5cm}{10.cm}{\includegraphics{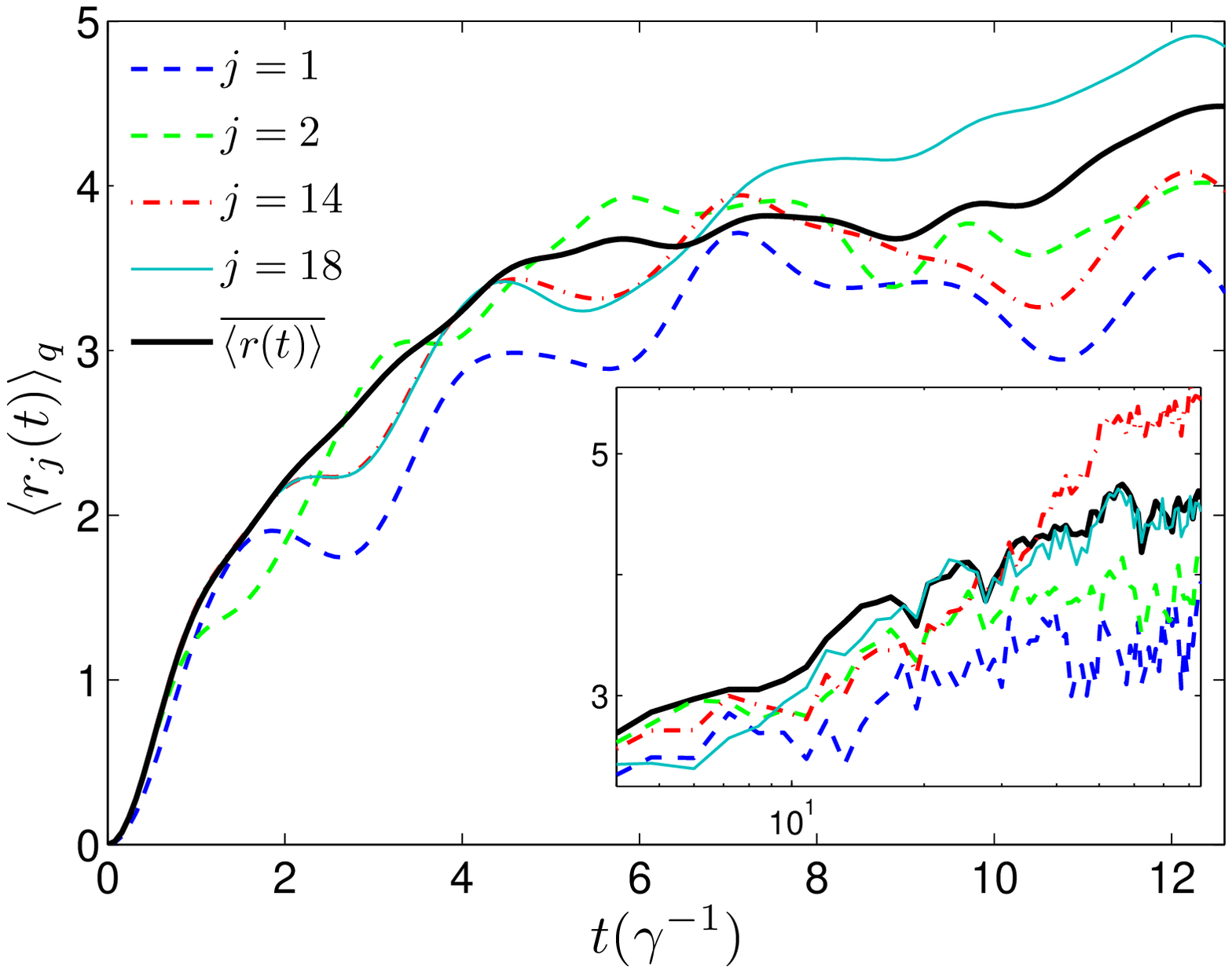}}
\caption{Average displacement $\langle r_j(t) \rangle_q$ for a
quantum walker which started at the $j$-th site on a DSG of
generation $g=5$. The main figure focuses on short times, while in the inset a wider temporal range is considered. Different colours and thicknesses distinguish
different initial sites $j$, as shown in the legend; the labeling
is the same as in Fig.~\ref{fig:pi_apex}. Notice the appearance of
local minima from $t \approx \gamma^{-1}$ onwards.}
\label{fig:r_DS}
\end{center}
\end{figure}

The dynamics of quantum particles on non-regular structures has
been investigated in several works meant to analyze the
quantum dynamics of tight-biding electrons in quasicrystals, in
aperiodic and quasi-periodic chains and in random environments
\cite{yin,yuan,grimm}. There, the highlighted dramatic deviations
from the ballistic behaviour (expected for regular, infinite
lattices) range from anomalous to superdiffusion, to
decoherence, and even to Anderson localization \cite{anderson}.

Here, we consider the case in which non-regularity stems from the intrinsic spatial inhomogeneity of the substrate (for DSGs and CTs) and we
also study quantum walks on STs which allow to evidence the role
of finiteness. From the experimental side, the importance of such factors (spatial inhomogeneities and finiteness of the sample) has been increasingly
recognized (see e.g. \cite{monastyrsky}), so that it is of great interest to understand to
what extent quantum transport is influenced by them.

First of all, we notice that the fact that $\pi_{k,j}(t)$ does not attain a stationary
distribution also causes the average displacement $\langle r_j(t)
\rangle_q$ not to necessarily increase monotonically
with $t$. Moreover, due to reflection effects, we expect the mean value $\overline{\langle r(t) \rangle}_q$
to overestimate the displacement performed by a CTQW that
started from a peripheral site. Indeed, one finds for the DSG and the CT
that  $\overline{\langle r(t) \rangle}_q$ is larger than $\langle r_{j}(t)
\rangle_q$, $j$ being  any corner of the gasket (see Fig.~\ref{fig:r_DS})
or any peripheral site, respectively.
As for the ST, $\overline{\langle r(t) \rangle}_q$ trivially
equals $\langle r_j(t) \rangle_q$, for all $j$.

\begin{figure}
\begin{center}
\resizebox{13.5cm}{10.5cm}{\includegraphics{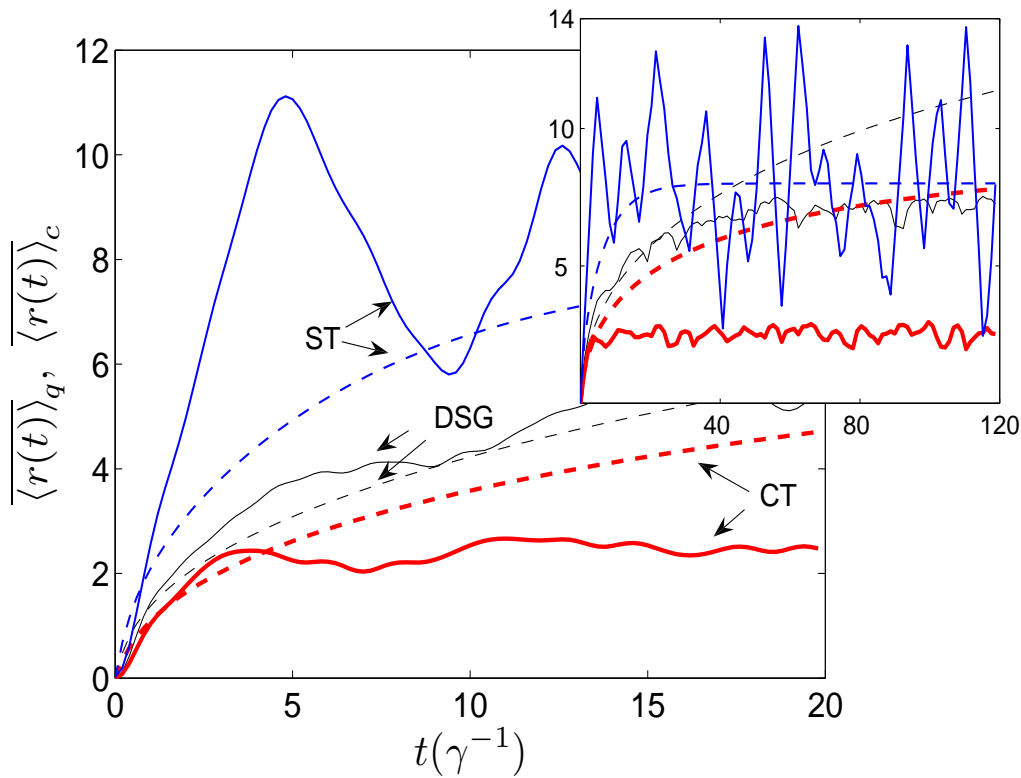}}
\end{center}
\caption{Classical (dashed line) and quantum (continuous line)
average displacement for the DSG ($g=5$, $\mathcal{N}=243$), the CT ($g=6$, $\mathcal{N}=190$) and the
ST ($L=16$, $\mathcal{N}=256$). The main figure focuses on the short-time regime,
while the inset also shows the long-time regime.}
\label{fig:r_av2}
\end{figure}

As mentioned above, for classical diffusion the average
displacement grows continuously from zero to a maximum value $r_c$
which, due to equipartition, is just the mean distance among
sites:
\begin{equation}\label{eq:r_c}
r_c = \frac{1}{\mathcal{N}^2}\sum_{k,j=1}^{\mathcal{N}} \ell(k,j).
\end{equation}
Despite of the oscillating behaviour of $\overline{\langle r(t)
\rangle}_q$, we can obtain an analogous constant value $r_q$,
around which the average displacement eventually fluctuates, which reads
\begin{equation}
r_q \equiv \lim_{T \rightarrow \infty} \frac{1}{T}  \int_0^T  dt \; \overline{\langle r(t) \rangle}_q .
\end{equation}
Otherwise stated, $\overline{\langle r(t) \rangle}_q$ eventually reaches a
``stationary regime'' in which it fluctuates around a constant value (see Fig.~\ref{fig:r_av2}).

Of course, $r_q$ and $r_c$ depend on both
the topology and the size of the substrate, and they diverge as
$\mathcal{N} \rightarrow \infty$. From the remarks of
Sec.~\ref{subsec:TP}, we expect that quantum
interference arising from reflection affects $r_q$, making it
smaller than $r_c$. Indeed, for the CT and the DSG, Fig.~\ref{fig:r_av2}
clearly shows that $r_q < r_c$; this is especially apparent for the CT
where $r_c \approx 8.9$ (calculated from Eq.~\ref{eq:r_c}) is nearly four
times larger than $r_q \approx 2.4$. Conversely, for the ST,
where interference only stems from the superposition of waves
which have crossed the whole substrate, we find that
$r_c$ and $r_q$ are eventually comparable. Therefore, we expect that on
structures endowed with reflecting boundaries (i.e. peripheral
nodes of low connectivity), at sufficiently long times, the
expectation value of the average distance reached by a quantum
particle is strictly smaller than the average distance $r_c$ among
the nodes.

\begin{figure}
\begin{center}
\resizebox{12.5cm}{10cm}{\includegraphics{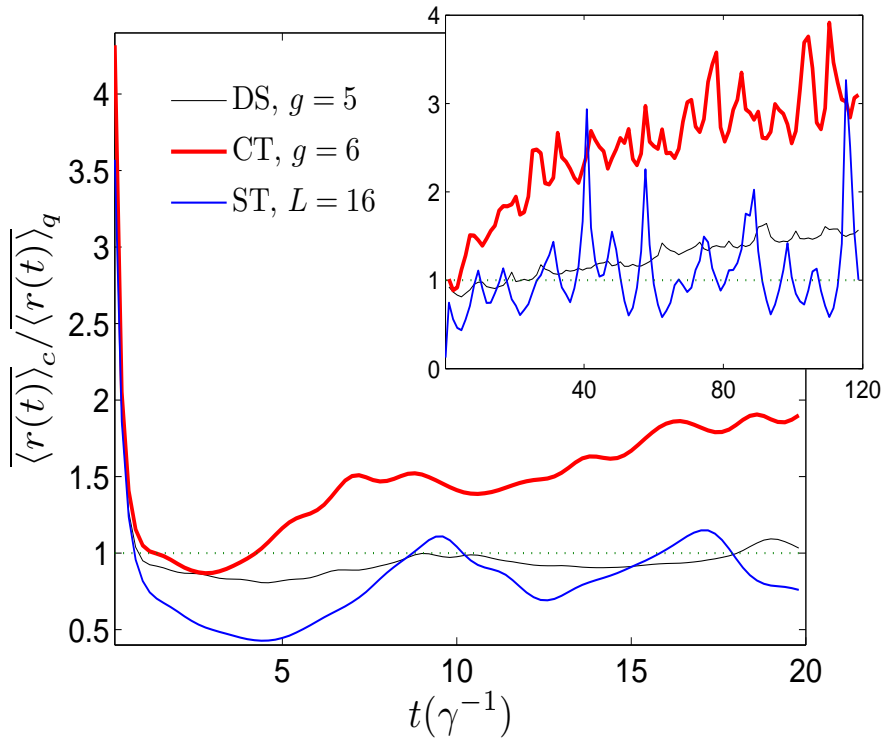}}
\end{center}
\caption{Ratio between classical and quantum average displacement
for the DSG ($g=5$), the CT ($g=6$) and the ST ($L=16$), as shown
by the legend. The main figure focuses on the short-time, while
the inset also shows the long-time regime. In particular, for the square torus, the ratio $\overline{\langle r(t)
\rangle}_c / \overline{\langle r(t) \rangle}_q$ is at first smaller than unity
and then it oscillates around unity; the highest peaks are signs of
(partial) revivals.} \label{fig:r_av}
\end{figure}

%Let us now focus on shorter times and compare how fast $\overline{\langle r(t) \rangle}_q$ and $\overline{\langle r(t) \rangle}_c$ reach the stationary state and the asymptotic value $r_c$, respectively. Referring to Fig.~\ref{fig:r_av2}, we can infer that after a very short initial period $\overline{\langle r(t) \rangle}_q$ grows faster than $\overline{\langle r(t) \rangle}_c$, but this speed-up of $\overline{\langle r(t) \rangle}_q$ is limited to relatively short times. More precisely, for the ST the speed-up is quadratic ($\overline{\langle r(t) \rangle}_q \sim t$) and it holds over a temporal range ($t < 5 \gamma ^{-1}$) wider than those found for the DSG and the CT. The latter structures also display a weaker, less-than-quadratic, speed-up; for the DSG we find that at short times ($t < 4 \gamma ^{-1}$), $\overline{\langle r(t) \rangle}_q$ scales with time according to a power a law with an exponent $\approx 0.6$, to be compared with  $1/d_w \approx 0.43$ (see Eq.~\ref{eq:anomalous_diffusion}). For the CT the stationary state is reached at even shorter times ($t < 2 \gamma ^{-1}$).

The classical and the quantum cases are further compared in Fig.~\ref{fig:r_av} which shows the ratio $\overline{\langle r(t) \rangle}_c / \overline{\langle r(t) \rangle}_q$: one can notice that, for significant times ($t>1 \gamma^{-1}$), the classical average displacement is strictly lower than the quantum-mechanical one up to time $t \approx 4 \gamma^{-1}, t \approx 9 \gamma^{-1}$ and $t \approx 18 \gamma^{-1}$ for CT, ST and DSG, respectively.

Thus, we can conclude that, on restricted geometries such as those
analyzed here, CTQWs can spread faster than their classical counterpart, although the advantage is significant only at relatively short times. Moreover, the spatial homogeneity enhances the speed-up; especially
for CTs and, in general, for tree-like structures, the large number of peripheral sites gives rise to localization effects which significantly reduce $r_q$.

Finally, we stress that
analytical results on the average displacement performed by a quantum particle on discrete structures are rather sparse (see e.g. \cite{vidal,katsanos}); in the Appendix we prove that on infinite $d$-dimensional hypercubic lattices both the average chemical displacement defined in Sec.~\ref{ssec:AD} and the Euclidean displacement depend linearly on time and that this kind of behaviour survives, at short times, also for finite lattices.

\subsection{Average return probability}
The average displacement for CTQWs already highlighted some
aspects of the role of inhomogeneities for transport processes. Now, we
obtain further insights by considering the average return probability.

For the DSG we can get $\bar{p}(t)$ without numerically
diagonalizing $\mathbf{L}$, since it only depends on eigenvalues
which can be calculated iteratively.
\begin{figure}
\begin{center}
\resizebox{12.5cm}{11cm}{\includegraphics{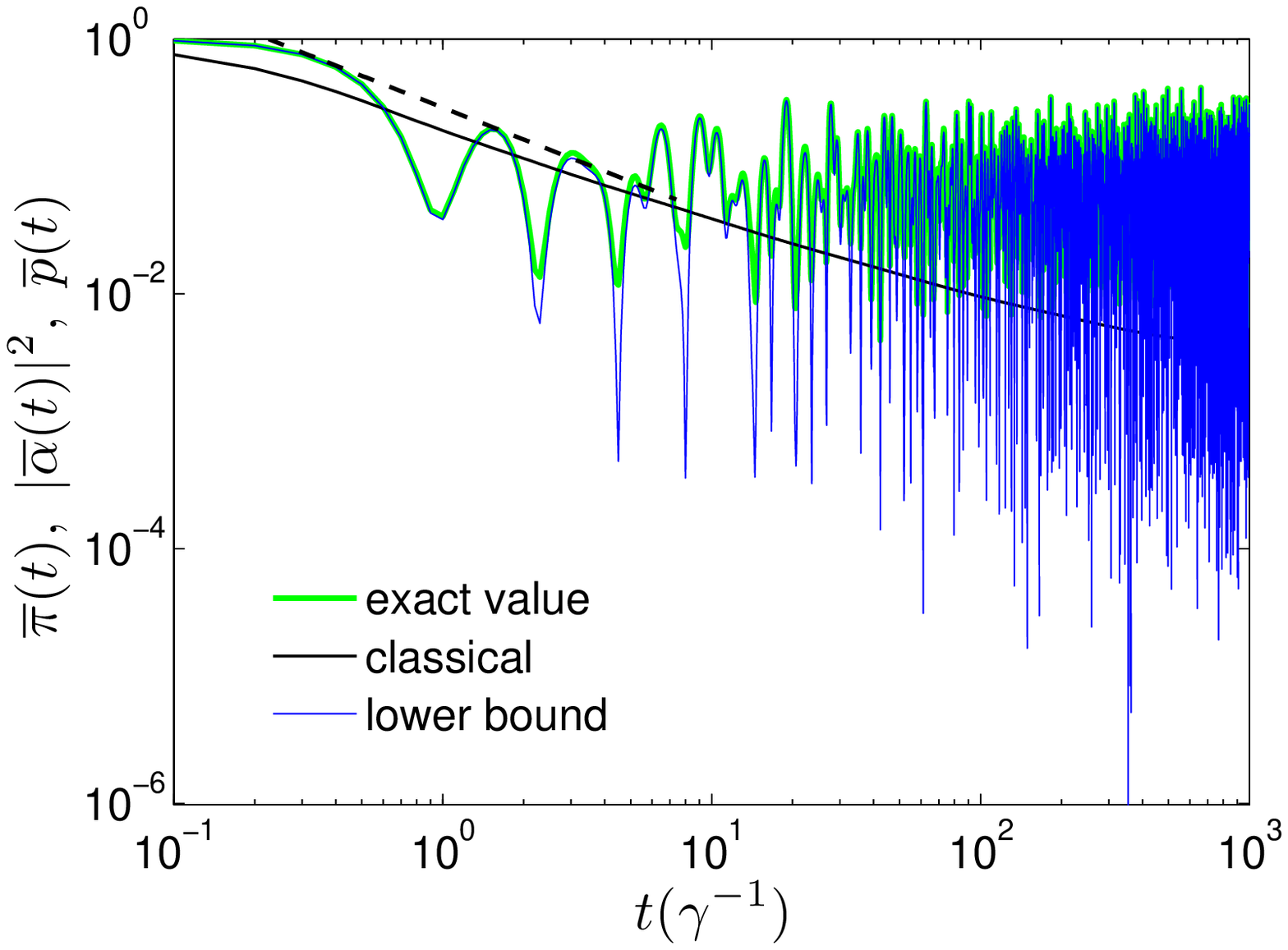}}
\end{center}
\caption{Average return probability $\bar{\pi}(t)$ for the DSG of generation $g=5$ on a log-log scale. The
comparison with the classical $\bar{p}(t)$ evidences that the
classical random walk spreads more efficiently than its quantum-mechanical
counterpart. The dashed line represents the envelope of $\bar{\pi}(t)$.} \label{fig:pi_compare}
\end{figure}
Figure \ref{fig:pi_compare} displays the averaged probabilities
$\bar{p}(t)$, $\bar{\pi}(t)$ and $|\bar{\alpha}(t)|^2$ -
numerically evaluated from Eqs.~\ref{eq:p_bar}, \ref{eq:pi_bar}
and \ref{eq:mu}, respectively - as a function of time, obtained
for $g=5$. The classical $\bar{p}(t)$ decays monotonically to the
equipartition value $1 / \mathcal{N}$, while the
quantum-mechanical probabilities eventually oscillate around the
value $0.7$, which is larger than $3^{-g}$. Although the
amplitude of fluctuations exhibited by the lower bound is larger
than that of the exact value, the agreement between the two
quantities is very good. In particular, the positions of the
extrema practically coincide and the maxima of
$\bar{\pi}(t)$ are well reproduced by the lower bound. An
analogous behaviour was found also for other graphs, such as
square lattices \cite{volta}, Cayley trees \cite{bierbaum} and
stars \cite{muelken}. Notice, however, that for the square
lattices the lower bound turns out to be exact while for Cayley
trees and for stars it is only an approximation, which, moreover,
turns out to be less accurate than what we find here for the DSG.

On short times ($t < 5 \gamma^{-1}$) it is possible to construct the
envelope of $\bar{\pi}(t)$, which depends algebraically on $t$. The
exponent is $\approx -0.82$, to be possibly compared with $\tilde{d}/2
\approx -0.68$ which is the exponent expected classically for the infinite
DSG. The decay of the average return probability $\bar{\pi}(t)$ for the ST
can be estimated as well: its envelope goes like $t^{-2}$ (classically as
$\bar{p}(t) \sim t^{-1}$)\cite{volta,muelken3}, implying a faster
delocalization of the CTQW over the graph.

\begin{center}
\begin{figure}
\resizebox{15.5cm}{10.5cm}{\includegraphics{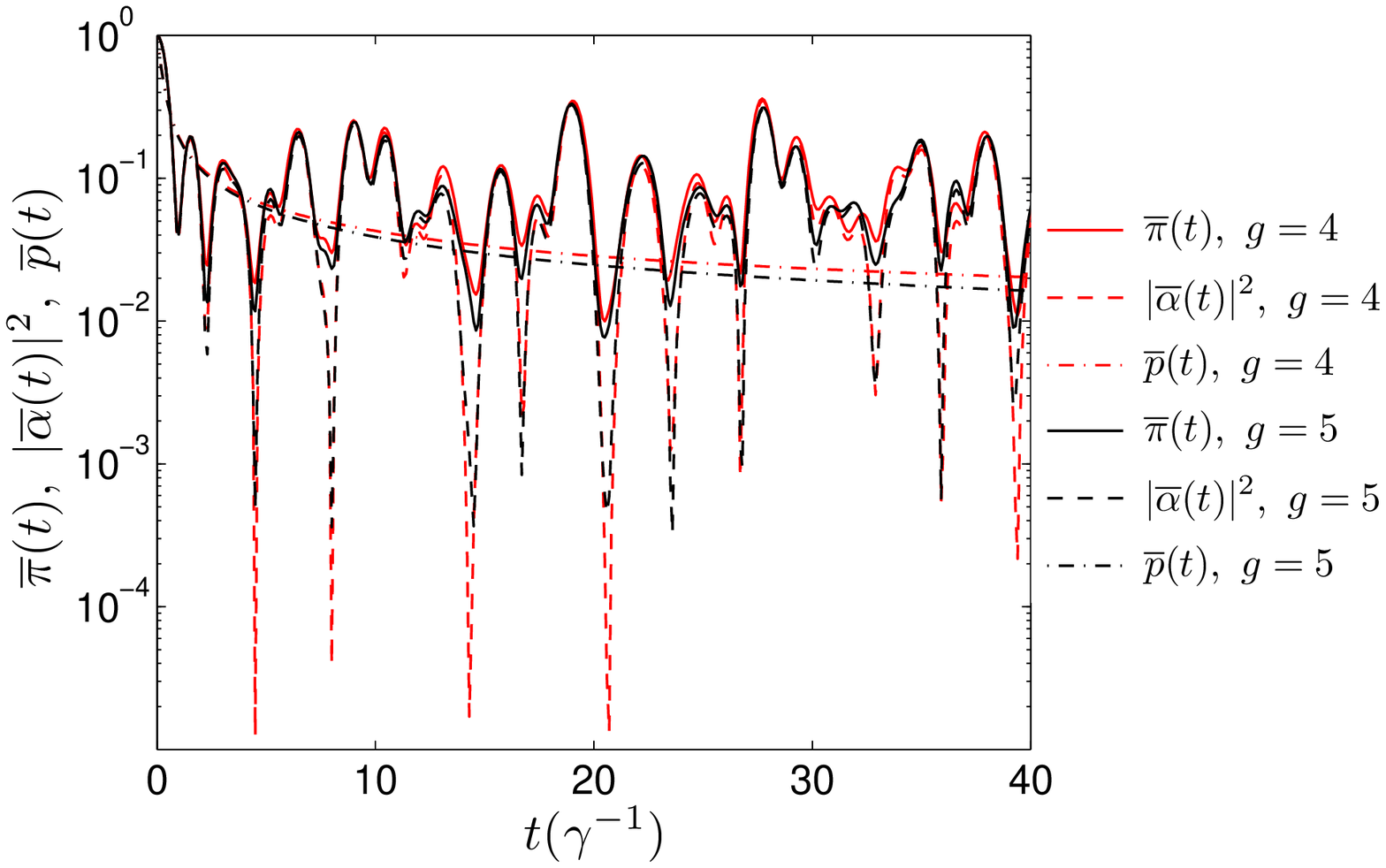}}
\caption{Average return probability $\bar{\pi}(t)$ for the DSG of generation $g=4$ (bright colour) and $g=5$ (dark colour). Its lower bound $|\bar{\alpha}(t)|^2$ (dashed line)
and the classical $\bar{p}(t)$ (dotted line) are also depicted, as shown by the
legend.} \label{fig:pi_G4G5}
\end{figure}
\end{center}

Interestingly, for the DSG, the overall shape of $\bar{\pi}(t)$ does not depend
significantly on the size of the gasket (Fig.~\ref{fig:pi_G4G5}).
In fact, the behaviour of $\bar{\pi}(t)$ is mainly controlled by
the most highly degenerate eigenvalues. These do not change when
increasing the fractal size (i.e. its generation). These values
are: $3$ with degeneracy $m_g(3)=(3^{g-1}+3)/2$, $5$ with
degeneracy $m_g(5)=(3^{g-1}-1)/2$, and $(5 \pm \sqrt{13})/2$ with
degeneracy $m_{g-1}(3)$, see Sec.~\ref{sec:DSG}.

\subsection{Long time averages}\label{ssec:LongTime}
As underlined in Sec.~\ref{sec:CTRW}, the unitary time evolution does not allow a
definite long-time limit for $\pi_{k,j}(t)$. Then, in order to
obtain information about the overall spreading of quantum walks,
it is advantageous to use the long time average (LTA):
\begin{equation}
\chi_{k,j} \equiv \lim_{T \rightarrow \infty} \frac{1}{T}
\int_{0}^{T} dt \; \pi_{k,j}(t) = \sum_{n,m} \delta_{\lambda_n,\lambda_m} \langle
k | \psi_n \rangle \langle \psi_n | j \rangle \langle j | \psi_m
\rangle \langle \psi_m | k \rangle,
\end{equation}
where $\delta_{\lambda_n,\lambda_m}$ equals $1$ for $\lambda_n=\lambda_m$ and is zero otherwise. The LTA of
$\bar{\pi}(t)$ follows as
\begin{equation} \label{eq:chi_bar}
\bar{\chi} \equiv \lim_{T \rightarrow \infty} \frac{1}{T} \int_0^T
dt \; \bar{\pi}(t) = \frac{1}{\mathcal{N}} \sum_{n,m,j}
\delta_{\lambda_n,\lambda_m}|\langle j | \psi_n \rangle|^2 \;
|\langle j | \psi_m \rangle|^2,
\end{equation}
for which we obtain a lower bound which does not depend on the
eigenvectors \cite{muelken}:
\begin{equation} \label{eq:chi_bar_lb}
\bar{\chi} \geq \frac{1}{\mathcal{N}^2} \sum_{n,m}
\delta_{\lambda_n, \lambda_m} \equiv \bar{\chi}_{lb}.
\end{equation}

\begin{figure}
\begin{center}
\resizebox{14.0cm}{11.0cm}{\includegraphics{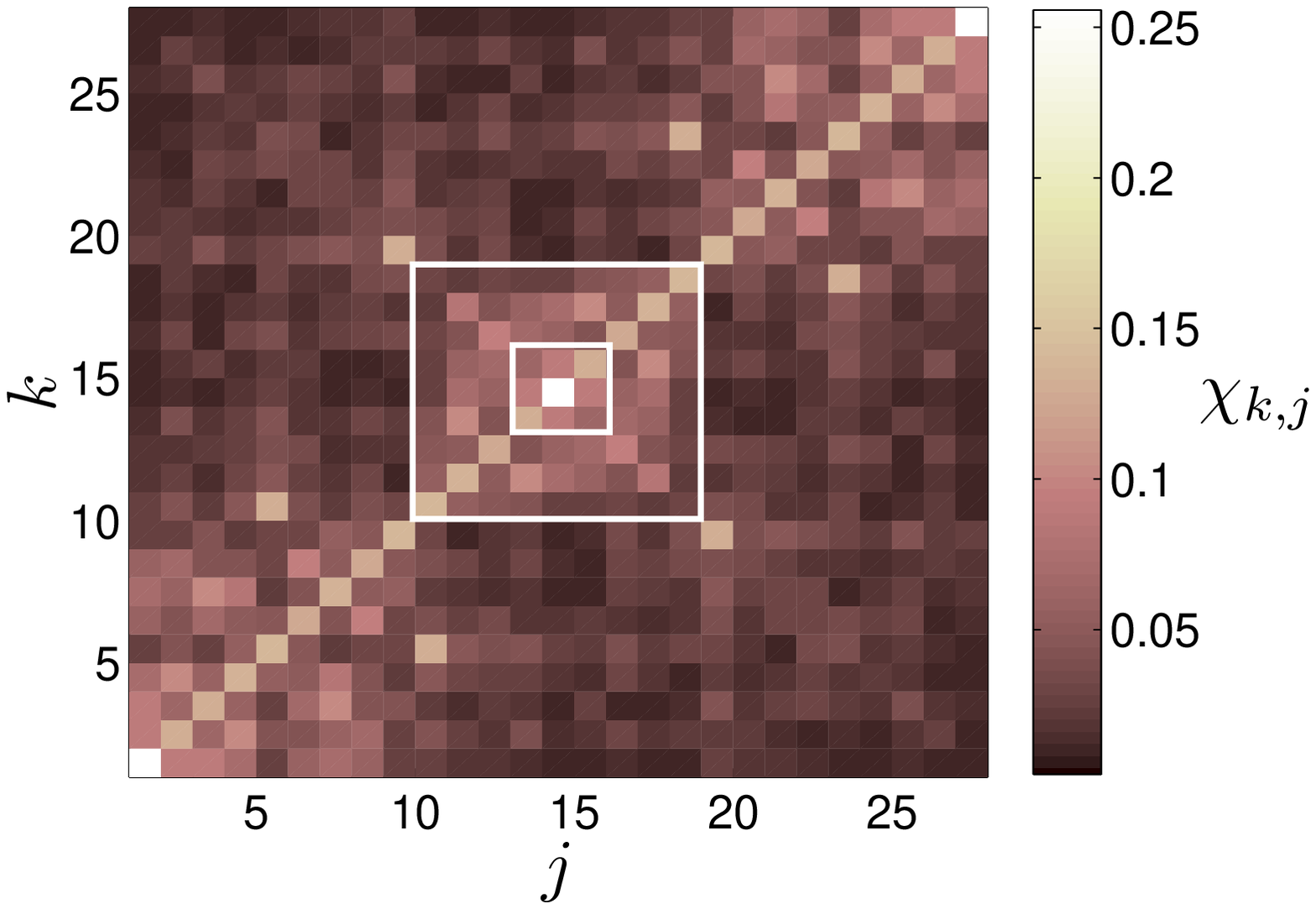}}
\resizebox{14.0cm}{11.0cm}{\includegraphics{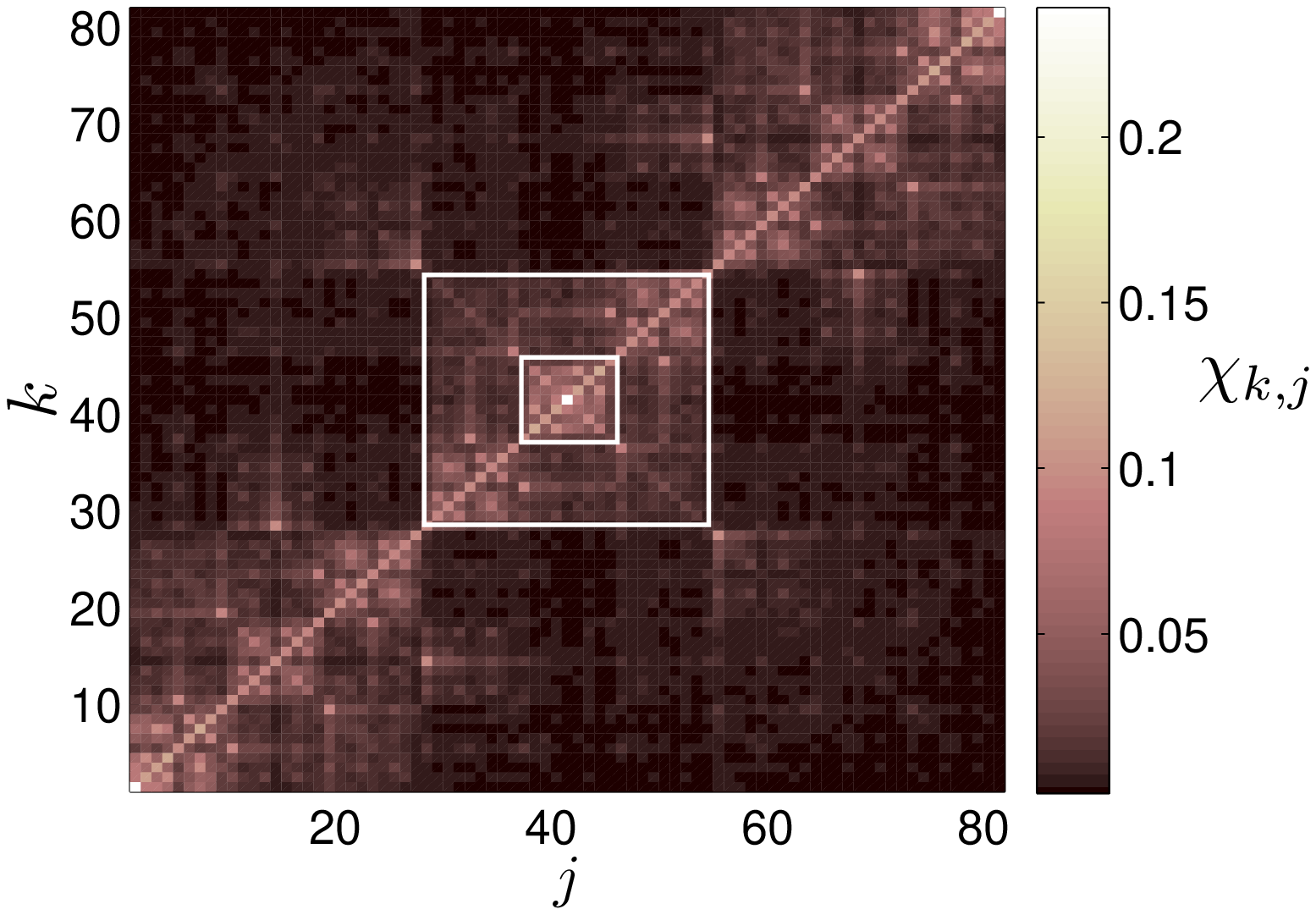}}
\caption{Limiting probabilities for the DSG of generation $g=3$
(top) and $g=4$ (bottom), whose volumes are $\mathcal{N}=27$ and
$\mathcal{N}=81$, respectively. The white lines enclose the
limiting distributions for gaskets of smaller generations. Notice
that the global maxima lay on the main diagonal and correspond to
$j=1,14,27$ and to $j=1,41,81$ for $g=3$ and for $g=4$,
respectively.} \label{fig:carpet}
\end{center}
\end{figure}

We first consider the DSG for which Fig.~\ref{fig:carpet} shows $\chi_{k,j}$ as a contour plot,
whose axes are labeled by the nodes $k=1,...,\mathcal{N}$ and
$j=1,...,\mathcal{N}$. Bright colours correspond to large, dark
ones to small LTAs. First of all, we notice that the LTAs are far
from being homogeneous and, hence, are not equipartioned. In
particular, the values on the main diagonal are high, meaning that
CTQWs have a high LTA probability to be at the starting node.

The inhomogeneity
of the pattern  mirrors the lack of translation invariance of the
DSG itself. For instance, $v$ being the label assigned to any
vertex of the main triangle, $\chi_{v,v}$ is a global maximum; off-diagonal local maxima correspond to couples of connected nodes belonging to different minor triangles of generation $g-1$. This allows to establish a mapping between the
pattern of $\chi_{k,j}$ and the structure of the relevant DSG.
Indeed, as suggested by the white delimiting lines in Fig.~\ref{fig:carpet},
the patterns of the LTA distributions exhibit self-similarity.

As for $\bar{\chi}$ and its lower bound
$\bar{\chi}_{lb}$, we recall that the former can be calculated numerically, once
all eigenvalues and eigenvectors of the Laplacian operator are
known (Eq.~\ref{eq:chi_bar}), while for the latter the knowledge of the
eigenvalue spectrum is sufficient (Eq.~\ref{eq:chi_bar_lb}).
Since the spectrum of the DSG is known, we
can calculate $\bar{\chi}_{lb}$ analytically.
%We start from Eq.~\ref{eq:chi_bar_lb}, where the sum runs over the set of all $\mathcal{N}$ eigenvalues, and we rewrite it in the ``space of distinct eigenvalues'' and then in the ``space of distinct degeneracies''.
Recalling the results of Sec.~\ref{sec:DSG}, at generation $g$ the spectrum of $\mathbf{L}$ displays $\tilde{\mathcal{N}}$ distinct eigenvalues, where
$$
\tilde{\mathcal{N}} = \sum_{r=0}^{g-1} 2^r + \sum_{r=0}^{g-2} 2^r + 1 = 3\times 2^{g-1}-1.
$$
We call the set of distinct eigenvalues $\{ \tilde{\lambda}_i \}_{i=1,...,\tilde{\mathcal{N}}}$.
Being $m(\lambda_i)$ the degeneracy of the eigenvalue $\lambda_i$, we can write
$$ \mathcal{N}^2 \bar{\chi}_{lb} = \sum_{n,m=1}^{\mathcal{N}}
\delta_{\lambda_n, \lambda_m} = \sum_{n=1}^{\mathcal{N}}
m(\lambda_n) = \sum_{i=1}^{\tilde{\mathcal{N}}} \left[ m(\tilde{\lambda}_i) \right]^2.
$$
Now, we go over to the space of distinct degeneracies, each
corresponding to a number $\rho$ of distinct eigenvalues and we
get the final, explicit formula
\begin{eqnarray} \label{eq:exact_chi_bar_lb}
\bar{\chi} & \geq & \bar{\chi}_{lb} = \frac{1}{\mathcal{N}^2} \sum_{r=0}^{2g} [m(r)]^2
\rho(m(r))
\nonumber \\
& & = \frac{1}{\mathcal{N}^2}  \left \{  \sum_{r=0}^{g-1} \left[
\frac{3^{g-r-1}+3}{2} \right]^2 \times 2^r +  \sum_{r=0}^{g-2}
\left[ \frac{3^{g-r-1}-1}{2} \right]^2 \times 2^r  +1  \right \}
\nonumber \\
& & = \frac{1}{3^{2g}} \left[ 3^g \left( 1 + \frac{3^g}{14}
\right) +\frac{10}{7}2^g - \frac{3}{2} \right] > \frac{1}{3^g}.
\end{eqnarray}
Interestingly, in the limit $g \rightarrow
\infty$, the LTA $\bar{\chi}$ is finite:
$$
\bar{\chi} \geq \lim_{g \rightarrow \infty} \bar{\chi}_{lb}
= \frac{1}{14},
$$
and $\bar{\chi}_{lb}$ reaches this asymptotic value from above.

\begin{figure}
\resizebox{16cm}{10cm}{\includegraphics{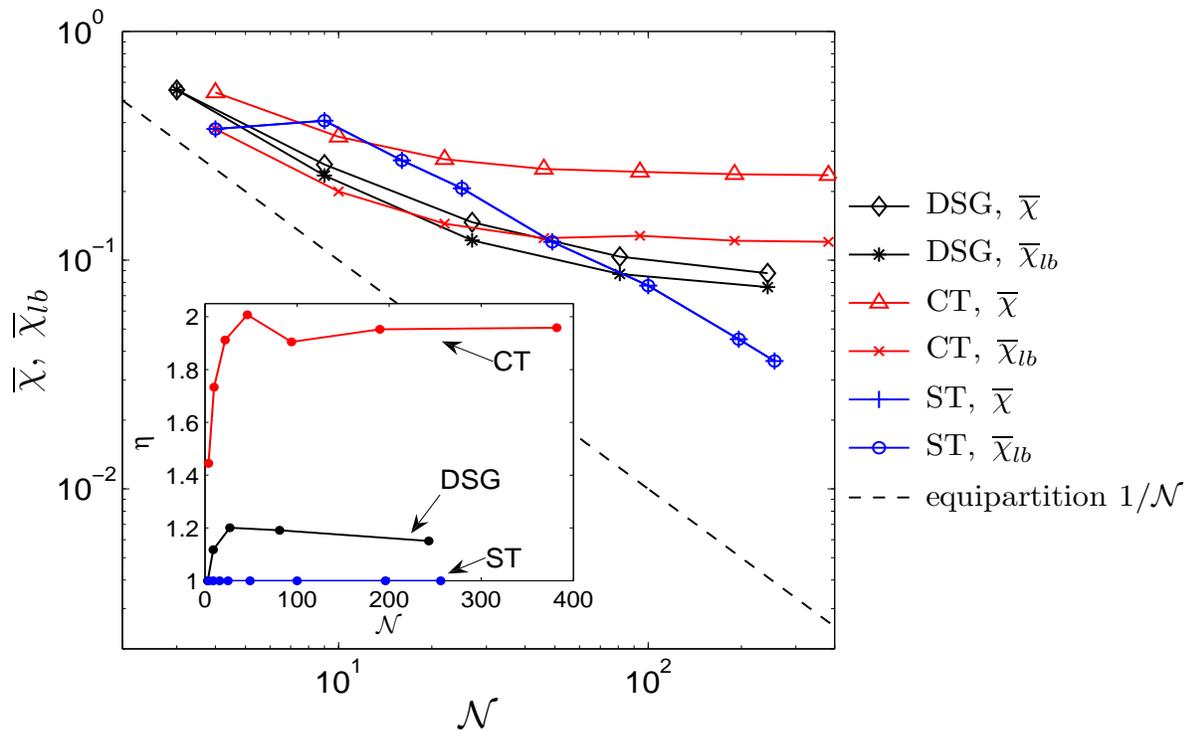}}
\caption{Long time average $\bar{\chi}$ and its lower bound
$\bar{\chi}_{lb}$ calculated according to
Eq.~\ref{eq:exact_chi_bar_lb} for the Dual Sierpinski Gasket (black
diamonds and stars), the Cayley Tree (red triangles and
crosses) and the Square Torus (blue crosses and open circles) versus the
size of the structure $\mathcal{N}$; lines are guides for the eye.
The dashed line represents the equipartition value
$1/\mathcal{N}$. Inset: ratio $\eta$ as a function of
$\mathcal{N}$. Notice that $\eta(\mathcal{N}) \equiv 1$ holds not
only for the periodic square lattice, but for all hypercubic lattices.} \label{fig:chi_bar}
\end{figure}

In Fig.~\ref{fig:chi_bar} we show, as functions of
$\mathcal{N}$, $\bar{\chi}$ and its lower bound, calculated
from Eq.~\ref{eq:chi_bar} and from
Eq.~\ref{eq:exact_chi_bar_lb}. For comparison, the same quantities obtained for
CTs and STs are also depicted. In the latter case, due to the regularity and
periodicity of the lattice \cite{bierbaum2}, the lower bound actually coincides with the exact value. For all cases
considered, $\bar{\chi}$ is larger than the equipartion value (given by the dashed line).

The inset of Fig.~\ref{fig:chi_bar} shows the ratio
$$
\eta(\mathcal{N}) \equiv \frac{\bar{\chi}}{\bar{\chi}_{lb}}.
$$
Obviously, the closer $\eta$ is to $1$, the better $\bar{\chi}_{lb}$
approximates $\bar{\chi}$. In this sense, the lower bound calculated for
CTs is not as good an approximation to $\bar{\chi}$ as it is for the DSG
and for the ST. For the CT, $\bar{\chi}_{lb}$ definitely
underestimates, being about half the exact value of $\bar{\chi}$. The quantity
$\eta(\mathcal{N})$ may act as
a measure of the inhomogeneity of a given substrate.  Practically, when
dealing with a large sized, sufficiently regular structure, we can get
information about the localization of a quantum particle moving on it
simply through $\bar{\chi}_{lb}$, thus avoiding the (lengthy) evaluation
of the eigenvector set.

\section{Conclusions}\label{sec:concl}
We investigated the behaviour of continuous-time quantum walks (CTQWs) on finite discrete structures characterized by different topologies; we considered the Square Torus (ST), the Cayley Tree (CT) and the Dual Sierpinski Gasket (DSG).

The interplay between the quantum-walk dynamics and the underlying topology was deepened by studying, in particular, the temporal evolution of the transfer probability distribution and the ratio $\bar{\chi} / \bar{\chi}_{lb}$ as a function of the substrate size. The latter turns out to be significantly sensitive to the inhomogeneity of the substrate, from which we can infer that lower bound estimates are especially reliable for regular structures.

From an applied, as well as theoretical, perspective, the average
displacement of the walker, as a function of time, also plays an important
role. This quantity is not only directly related to the transport
properties, but it also provides information about how fast the walk
explores the underlying structure, allowing an immediate comparison with
the classical case. We found that at short times, CTQWs can spread
faster than their classical counterparts, although spatial inhomogeneities and
finiteness jointly reduce this effect. In the Appendix we prove that for infinite $d$-dimensional hypercubic lattices, at long times both the average chemical and the Euclidean displacements depend
linearly on time (i.e. the motion is ballistic); for finite lattices this kind of behaviour holds at relatively short times only.

\section*{Acknowledgements}
EA thanks the Italian Foundation ``Angelo della Riccia'' for
financial support. Support from the Deutsche Forschungsgemeinschaft (DFG), the Fonds
der Chemischen Industrie and the Ministry of Science, Research and
the Arts of Baden-W\"{u}rttemberg (AZ: 24-7532.23-11-11/1) is
gratefully acknowledged.

\appendix
\section{Average chemical displacement on hypercubic lattices}
Here we consider infinite $d$-dimensional hypercubic lattices and, by exploiting their translational invariance, we prove that on them the average chemical displacements of CTQWs, as defined in
Sec.~\ref{ssec:AD}, depend linearly on time. We first focus on the
infinite discrete chain, then we consider the generic $d$-dimensional case and finally we analyze the two-dimensional lattice.

For a ring of length $\mathcal{N}$, by exploiting the Bloch states, we have \cite{muelken2}
\begin{equation}
\alpha_{k,j}(t) = \frac{1}{\sqrt{\mathcal{N}}} \sum_{l} e^{-i \lambda_l t} e^{-i l (k-j)},
\end{equation}
where $\lambda_l$ is the $l$-th eigenvalue of the Laplacian matrix $\mathbf{L}$ associated with the ring. In the limit $\mathcal{N} \rightarrow \infty$ we are allowed to replace the sum over $l$ by an integral, obtaining
\begin{equation}
\lim_{\mathcal{N} \rightarrow \infty} \alpha_{k,j}(t) = i^{k-j} e^{ - i 2 t } J_{k-j}(2t),
\end{equation}
where $J_k(z)$ is the Bessel function of the first kind. In the calculation of the transfer probability
$\pi_{k,j}(t)$ the phase factor vanishes and we have $\pi_{k,j}(t) = J_{k-j}^2(2t)$, which can be restated as
\begin{equation} \label{eq:probability}
\pi_{k,0}(t) = J_{k}^2(2t),
\end{equation}
due to the translational invariance of the structure. Clearly (in agreement with $\pi_{k,0}(t)$ being a probability distribution), one has for all $t$
\begin{equation} \label{eq:normalization}
\sum_{k=-\infty}^{+\infty} \pi_{k,0}(t) = \sum_{k=-\infty}^{+\infty} J_{k}^2(2t) =  J_0^2(2t) + 2 \sum_{k=1}^{\infty} J_{k}^2(2t) =1,
\end{equation}
the last equality being based on $J_k(z)=(-1)^k J_{-k}(z)$ and on Eq.~8.536.3 in \cite{integral}.

Now, the average chemical displacement of a CTQW which starts from
$0$ and moves on an infinite chain (subscript $q,1$) follows from
Eq.~\ref{eq:exploration_depth} as
\begin{eqnarray} \label{eq:Bessel_1}
\langle r_0 (t) \rangle_{q,1} & = & \langle r(t) \rangle_{q,1}  =
\sum_{k \in V} \ell(k,0) \, J_{k}^2(2t)   \\
\nonumber
& & = \sum_{k = -\infty}^{\infty} |k| \, J_{k}^2(2t)
= 2 \sum_{k = 1}^{\infty} k \, J_{k}^2(2t).
\end{eqnarray}
Here, in the first equality we dropped the subscript $0$ due to the equivalence between the
sites and in the last equality we exploited the symmetry of the Bessel functions, $ J_{-k}^2 (z)  = J_{k}^2 (z)$.
Now, recalling the recursion formula Eq.~8.471.1 in \cite{integral}
\begin{equation}\label{eq:Bessel_rec}
J_{k-1}(z) + J_{k+1}(z) = \frac{2k}{z} J_k(z),
\end{equation}
we can write
\begin{equation}\label{eq:Bessel_2}
2 \sum_{k = 1}^{\infty} k \, J_{k}^2(z) = z \sum_{k = 1}^{\infty} [J_{k-1}(z) \, J_{k}(z) + J_{k}(z) \, J_{k+1}(z)] \equiv z \mathcal{J}(z),
\end{equation}
by defining the function $\mathcal{J}(z)$. Hence
\begin{equation}\label{eq:Bessel_conc}
\langle r_k \left( \frac{z}{2} \right ) \rangle_{q,1} = z \mathcal{J}(z),
\end{equation}
where we put $2t=z$.
The squared Bessel function $J_k^2(z)$ is almost everywhere positive
%($k \in \mathbb{R}^{+}$ and the zeros of the Bessel function are real and countable)
and the analysis of its zeros allows to state that, for $t>0$, the sum appearing in the left-hand-side of Eq.~\ref{eq:Bessel_conc} is strictly positive; the same holds therefore for $\mathcal{J}(z)$, for which we also notice from Eq.~\ref{eq:Bessel_2} that $\mathcal{J}(0)=0$.
%the sum in the right-hand side.
Moreover, through the following recursion formula, Eq.~8.471.2 in \cite{integral}
\begin{equation}\label{eq:Bessel_rec2}
2 \frac{\partial}{\partial z} J_k(z) = J_{k-1}(z) - J_{k+1}(z),
\end{equation}
it follows by directly differentiating $\mathcal{J}(z)$ and rearranging the terms
\begin{equation}\label{eq:Bessel_4}
\frac{d}{d z} \mathcal{J}(z) = \frac{J_0(z)[J_0(z) + J_2(z)]}{2} = \frac{J_0(z)J_1(z)}{z},
\end{equation}
where in the last expression we again used Eq.~\ref{eq:Bessel_rec} for $k=1$.
The indefinite integral of Eq.~\ref{eq:Bessel_4} is (see Eq.~5.53 in \cite{integral})
\begin{equation}\label{eq:indefinite_integral}
\mathcal{J}(z) = z \, J_0^2(z) + z \, J_1^2(z) - J_0(z)J_1(z) + C,
\end{equation}
as can be simply verified by differentiating Eq.~\ref{eq:indefinite_integral} and using Eqs.~\ref{eq:Bessel_rec} and \ref{eq:Bessel_rec2}. Furthermore, since $\mathcal{J}(0)=0$, we have $C=0$.

Therefore, the following, for us fundamental, relation holds:
\begin{equation}\label{eq:Bessel_3}
\sum_{k = 1}^{\infty} k \, J_{k}^2(z)
 = \frac{z}{2}[z \, J_0^2(z) + z \, J_1^2(z) - J_0(z)J_1(z)].
\end{equation}

Now, from Eqs.~\ref{eq:Bessel_1} and \ref{eq:Bessel_3} we get the exact expression for the average chemical displacement
\begin{equation}\label{eq:Bessel_1_fin}
\langle r \left( \frac{z}{2} \right) \rangle_{q,1} = z \left[ z \, J_0(z)^2 + z \, J_1(z)^2 - J_0(z)J_1(z) \right].
\end{equation}
%the time derivative of $\mathcal{J}(t)$ vanishes as
%$t \rightarrow \infty$, namely $\mathcal{J}(t)$ approaches a positive value independent of $t$. %Consequently, from Eqs.~\ref{eq:Bessel_1} and \ref{eq:Bessel_2} we infer that
%(one can check numerically that indeed it goes to $2/\pi$).
For large $z=2t$ (i.e. long times) we can use the expansion (see Eq.~8.451.1 in \cite{integral})
\begin{equation} \label{eq:Bessel_expansion}
J_k(z) = \sqrt{\frac{2}{\pi z}}\left[ \cos \left( z -  \frac{k\pi}{2} - \frac{\pi}{4} \right) + O \left( \frac{1}{z} \right) \right].
\end{equation}
%to get
%\begin{equation}
%\mathcal{J}(z) = \frac{2}{\pi} + \frac{\cos(2z)}{\pi z}  + \left[ \cos \left( z - \frac{\pi}{4} \right) - \sin \left( z - \frac{\pi}{4} \right) \right] O \left( \frac{1}{z} \right).
%\end{equation}
%\begin{equation}
%\mathcal{J}(z) = \frac{2}{\pi} + \frac{\cos(2z)}{\pi z}  + \frac{4 \sqrt{2}}{\pi} \sin z \; O \left( \frac{1}{z} \right).
%\end{equation}
Consequently, inserting Eq.~\ref{eq:Bessel_expansion} for $J_0(z)$ and $J_1(z)$ into \ref{eq:Bessel_1_fin}, we infer that the long time behaviour of the average chemical CTQW displacement on an infinite chain obeys
%\begin{equation}\label{eq:Bessel_1_last}
%\langle r(t) \rangle_{q,1} = 2t \left[ \frac{2}{\pi} + f(t) \; O(1/t) \right] \sim \frac{4t}{\pi},
%\end{equation}
\begin{equation}\label{eq:Bessel_1_last}
\langle r(t) \rangle_{q,1} \sim \frac{4t}{\pi}.
\end{equation}
%where $f(t)$ is a finite oscillating function;
This result is consistent with findings reported in \cite{katsanos} for the average square displacement.

\begin{figure}
\resizebox{12.5cm}{10cm}{\includegraphics{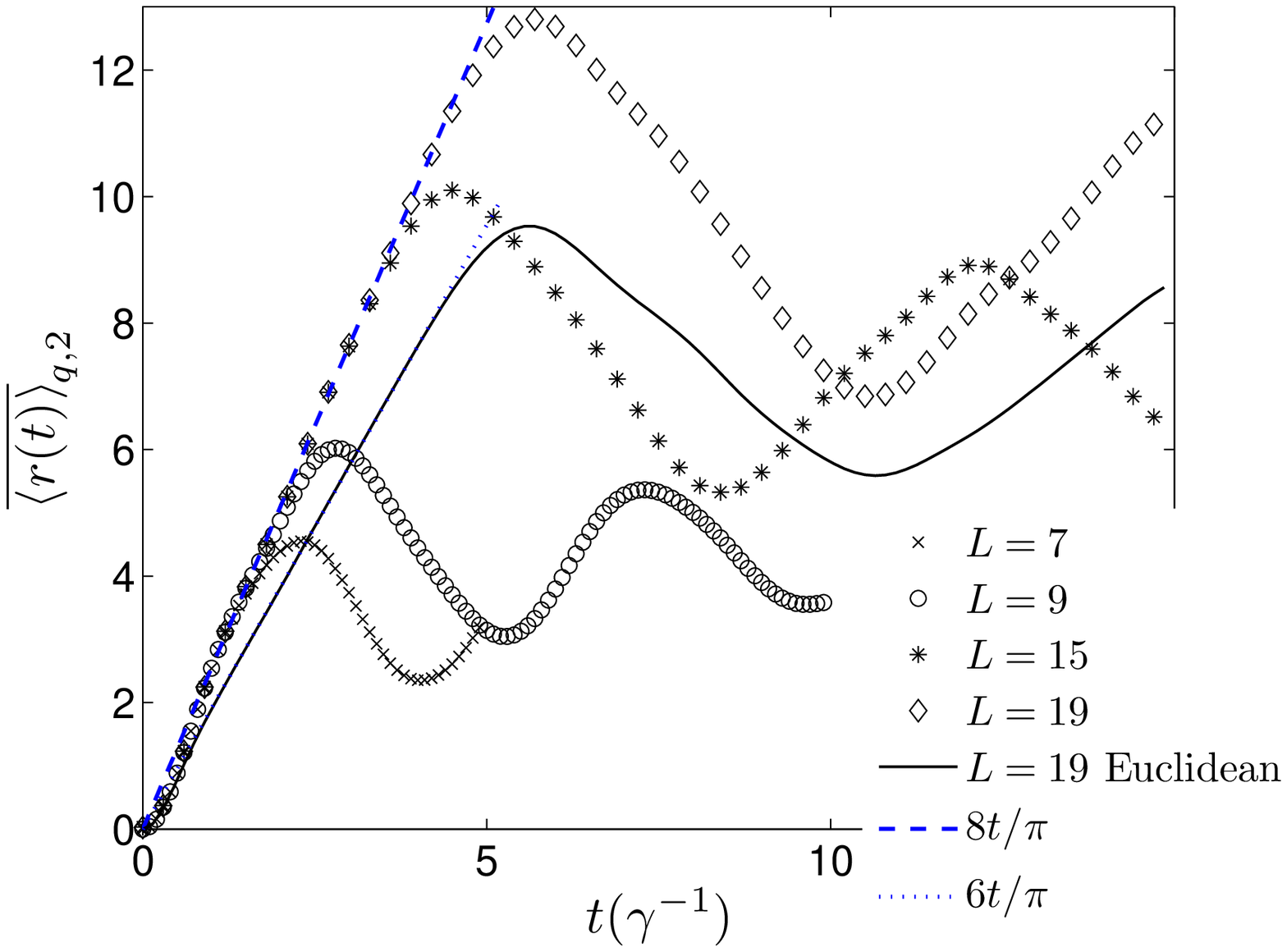}}
\caption{Average chemical displacement $\overline{\langle r (t) \rangle}_q$
for the torus  calculated according to
Eq.~\ref{eq:av_cover_length} (symbols); the line represents the Euclidean average displacement for $L=19$.
%and Eq.~\ref{eq_r_Bessel}
%(line).
%Different sizes are depicted, as shown by the legend.
The dotted and the dashed lines highlight the linear dependence on $t$ exhibited by the average chemical distance and by the average Euclidean distance, respectively.} \label{fig:bessel}
\end{figure}

Let us now consider higher dimensional hypercubic lattices. Again, without loss of generality, we can assume the CTQW to start from the point $\mathbf{0} = (0,0,...,0)$ so that the chemical distance attained by a walker being at the generic site $\mathbf{k}=(k_1, k_2, ..., k_d)$ is $\ell(\mathbf{k},\mathbf{0}) = |k_1| + |k_2| + ... + |k_d|$. Furthermore, on a hypercubic lattice, assuming symmetric conditions in all directions, the probability distribution $\pi_{\mathbf{k},\mathbf{0}}(t)$ factorizes into the $d$-independent one-dimensional distributions $\pi_{k_j,0}(t)$:
%According to Eq.~\ref{eq:exploration_depth}, in order to get the average chemical displacement $\langle r(t) \rangle_{q,d}$ we need to average $\ell({\mathbf{k},\mathbf{0}})$ with respect to the transfer probability which can be suitably written as
\begin{equation}\label{eq:Bessel_d}
\pi_{\mathbf{k},\mathbf{0}}(t)= \pi_{k_1,0}(t) \, \pi_{k_2,0}(t) \, ... \,  \pi_{k_d,0}(t) = \prod_{j=1}^{d} \pi_{k_j,0}(t).
\end{equation}
Hence
\begin{equation}
\langle r_{\mathbf{0}}(t) \rangle_{q,d} =  \langle \sum_{j=1}^d |k_j| \rangle_{q,d} = \sum_{j=1}^d  \langle |k_j| \rangle_{q,d} = \sum_{j=1}^d \langle |k_j| \rangle_{q,1} = d \langle r_0(t) \rangle_{q,1}.
\end{equation}
In the last relation we used the fact that $\langle |k_j| \rangle_{q,d} = \langle |k_j| \rangle_{q,1}$ since for each $j$ only the distribution $\pi_{k_j,0}(t)$ matters, the other distributions adding up to a factor of unity each.
%One therefore has
%\begin{equation} \label{eq:contributions}
%\langle r(t) \rangle_{q,d} = \sum_{\mathbf{k} \in V} \ell(\mathbf{k},\mathbf{0}) \prod_{j=1}^d \pi_{k_j,0}
%= \sum_{j=1}^d \sum_{k_j=1}^{\infty} k_j \pi_{k_j,0}
%\end{equation}
%where for the last equality we used the fact that the $\pi_{k_j,0}(t)$ are normalized and act on $d$ different $j$-subspaces.
%Equation \ref{eq:contributions} implies that $\langle r(t) \rangle_{q,d}$ separates into the sum of $d$ contributions:
%\begin{equation}
%\langle r(t) \rangle_{q,d} = \sum_{j=1}^d \langle k_j(t) \rangle,
%\end{equation}
%each contribution being equivalent (due to the instrinsic isotropy of the system) and equal to the average chemical displacement performed on the one-dimensional lattice, hence
Hence
\begin{equation}\label{eq:Bessel_d_fin}
\langle r(t) \rangle_{q,d} = d \, \langle r(t) \rangle_{q,1} \sim \frac{4dt}{\pi}.
\end{equation}
In particular, for the square lattice we have
\begin{equation}\label{eq:Bessel_d_fin}
\langle r(t) \rangle_{q,2} \sim \frac{8t}{\pi},
\end{equation}
which was used in Fig.~\ref{fig:bessel} (dashed line) to fit data relevant to the average chemical displacement performed by a CTQW on square tori of different (finite) sizes. As can be seen from the figure, the ballistic behaviour also holds for finite lattices, but for relatively short times only: at longer times the finiteness of the lattice starts to
matter and the product of Bessel functions in Eq.~\ref{eq:Bessel_d} ceases to be
a good approximation of the transfer probability. When the waves associated with
CTQWs have crossed the whole lattice, interference effects start to occur and
$\langle r(t) \rangle_{q,2}$ exhibits a non-monotonic behaviour. From the same figure we also notice that the $O(1/t)$ contributions of Eq.~\ref{eq:Bessel_expansion} get to be negligible for $t > 1 \, \gamma^{-1}$.

In Fig.~\ref{fig:bessel} we also show data for the average Euclidean displacement which displays a ballistic behaviour at short times as well. Indeed, for a hypercubic lattice of arbitrary dimension $d$, the following relation holds (see e.g. \cite{metric})
\begin{equation}
\frac{1}{\sqrt{3}} \ell(\mathbf{k},\mathbf{j}) \leq ||\mathbf{k} - \mathbf{j}|| \leq \ell(\mathbf{k},\mathbf{j})
\end{equation}
where $||\mathbf{k} - \mathbf{j}||$ denotes the Euclidean distance between the lattice points $\mathbf{k}$ and $\mathbf{j}$ chosen arbitrarily. By averaging each term of the previous equation with respect to the transfer probability $\pi_{\mathbf{k},\mathbf{j}}(t)$ (we can again exploit the translational invariance of the substrate and fix $\mathbf{j} = \mathbf{0}$), we find that the average Euclidean distance also scales linearly with time with a multiplicative factor bounded between $4 d/(\sqrt{3} \pi) \approx 2.31 d / \pi$ and $4d / \pi$. In particular, for the square torus of size $L=19$ considered in Fig.~\ref{fig:bessel}, we find that at relatively short times the average Euclidean distance scales as $6t / \pi$.

\end{document}